\documentclass{aa}

\usepackage{graphicx}
\usepackage{txfonts}
\usepackage{enumitem}
\usepackage[breaklinks, color links, citecolor=blue]{hyperref}
\usepackage{multirow}
\usepackage{natbib}
\usepackage[utf8]{inputenc}
\usepackage{gensymb} 
\usepackage{ulem} 
\usepackage{color, soul} 

\begin{document} 

   \title{Faraday tomography of LoTSS-DR2 data: \\III. Revealing the Local Bubble and the complex of local interstellar clouds in the high-latitude inner Galaxy}

   \titlerunning{Faraday tomography of LoTSS-DR2 data III.}
   \authorrunning{Erceg et al.}
  
   \author{Ana Erceg\inst{1,2}$^{\star}$,
           Vibor Jeli\'c\inst{1}$^{\star\star}$,
           Marijke Haverkorn \inst{2},
           Lovorka Gajovi\'c\inst{3},
           Martin Hardcastle\inst{4},
           Timothy W. Shimwell\inst{5,6},
           Cyril Tasse\inst{7,8}
          }

   \institute{Ru{\dj}er Bo\v{s}kovi\'c Institute, Bijeni\v{c}ka cesta 54, 10000 Zagreb, Croatia, \email{aerceg@irb.hr$^{\star}$; $^{\star\star}$vibor@irb.hr}
         \and 
            Department of Astrophysics/IMAPP, Radboud University, P.O. Box 9010, 6500 GL Nijmegen, The Netherlands
         \and 
            Hamburger Sternwarte, Universität Hamburg, Gojenbergsweg 112, 21029 Hamburg, Germany
         \and 
            Department of Physics, Astronomy and Mathematics, University of Hertfordshire, College Lane, Hatfield AL10 9AB, UK
        \and
             ASTRON, Netherlands Institute for Radio Astronomy, Oude Hoogeveensedijk 4, 7991 PD, Dwingeloo, The Netherlands
         \and
             Leiden Observatory, Leiden University, P.O. Box 9513, 2300 RA Leiden, The Netherlands
        \and
            GEPI \& ORN, Observatoire de Paris, Universit\'e PSL, CNRS, 5 Place Jules Janssen, 92190 Meudon, France
        \and
            Department of Physics \& Electronics, Rhodes University, PO Box 94, Grahamstown, 6140, South Africa
         }
   \date{Received 22 March, 2024 ; accepted 18 June, 2024}

  \abstract
   {The LOw-Frequency ARray (LOFAR) provides a unique opportunity to probe the magneto-ionised structure of our Galactic neighbourhood with great resolution. In this work, we present a new mosaic created with the second release of LOFAR Two-Metre Sky Survey data (LoTSS-DR2), which probes polarised synchrotron emission in the high-latitude inner Galaxy. This is the third paper in a series whose main goal is understanding the LOFAR Faraday tomographic data at low radio frequencies and utilising it to explore the intricate structure of the local interstellar medium (ISM).}
   {Our objective is to characterise the observed emission through multi-tracer analysis to better understand the volume and the structures that may be observed with LOFAR. Furthermore, we exploit Faraday depth as a unique tool to probe the diffuse magnetised structure in the local ISM. }
   {We produced a mosaic Faraday cube of LoTSS-DR2 data by applying a rotation measure synthesis algorithm. From the cube, we constructed Faraday moment maps to characterise the nature of spectra. Additionally, we quantified the linear depolarisation canals using the Rolling Hough transform and used them to search for alignment with other data sets. Utilising LoTSS-DR2 observations alongside complementary data sets including \textit{Planck} polarisation data, HI emission maps, and starlight polarisation measurements, we explored conditions along observed lines of sight and estimated the distance to the Faraday structures. }
   {The Faraday cube reveals a remarkably ordered structure across two-thirds of the observed area, whose orientation aligns well with that of both the HI filaments and the magnetic field. We estimate the minimum distance to the Faraday structures to be between 40 and 80 pc, which puts them in the vicinity of the Local Bubble wall. The emission is organised in a large gradient in Faraday depth whose origin we associate with the curved wall of the Local Bubble.}
   {Comparing our data with a model of the Local Bubble wall, we conclude that we might be probing a contribution of the medium inside the Local Bubble cavity as well, corresponding to the complex of local interstellar clouds. Moreover, we propose a toy model incorporating an ionised front of finite thickness into the Local Bubble wall, as a curved, cold neutral shell alone is insufficient to produce the observed gradient. We explore possible magnetic field strengths, as well as the possible distribution of the neutral and ionised medium inside the wall, within the constraints of the observed Faraday depth.
   }

   \keywords{ISM: general, structure, magnetic fields - radio continuum: ISM - techniques: polarimetric, interferometric, local interstellar matter}

   \maketitle

\section{Introduction}\label{sec:intro}
The LOw-Frequency Array (LOFAR) is a radio telescope operating at low radio frequencies, from 10 to 240 MHz. The LOFAR observations of our Galaxy probe the interaction between synchrotron emission and the ISM along the line of sight (LOS). Synchrotron emission is produced by ultra-relativistic cosmic ray electrons spiralling around magnetic field lines. Approximately 75\% of synchrotron emission exhibits linear polarization \citep{rybicki86}. As the emission passes through the magneto-ionic medium, the plane of polarization undergoes a rotation by an angle $\Delta \theta$. This effect is called Faraday rotation. The amount of rotation can be described using a quantity called Faraday depth, $\Phi$, which depends on physical conditions of the intervening ISM: electron density $n_e$, the LOS component of the magnetic field $B_\parallel$, and its extent along the LOS, $dl$. Faraday depth is defined as an integral
\begin{equation}\label{eq:FRangle}
     \frac{\Phi}{\mathrm{[rad \ m^{-2}]}} = 0.81 \int_{0}^{d} \frac{n_e}{\mathrm{[cm^{-3}]}} \frac{B_\parallel}{\mathrm{[\mu G]}} \frac{\mathrm{dl}}{\mathrm{[pc]}},
\end{equation}
evaluated from the source at 0 to the observer at a distance $d$ and can be positive or negative if the LOS magnetic field is oriented towards or away from us, respectively. In a simple scenario, when a background polarised source encounters a foreground Faraday rotating medium, the rotating and emitting volumes are separated and the cumulative change in polarisation angle along the LOS is linear with $\lambda^2$. The slope of this relation defines the rotation measure \citep[RM, e.g.][]{manchester72, ferriere21},
\begin{equation}\label{RMvslambda2}
    \frac{\Delta\theta}{\mathrm{[rad]}}=\frac{RM}{\mathrm{[rad \ m^{-2}]}}\frac{\lambda^2}{\mathrm{[m^2]}}.
\end{equation}
In this case, the Faraday depth of the observed polarised emission is equivalent to RM. However, most LOFAR observations show a more complex LOS distribution, characterised by mixing of the rotating and emitting volumes \citep{erceg22}.

Faraday rotation of each emitting layer is proportional to the square of the wavelength $\lambda^2$, which makes longer wavelengths (i.e., lower frequencies) more susceptible to the rotation. Therefore, low-frequency observations are capable of detecting very low-density ISM in a weak magnetic field. However, they are also more sensitive to depth depolarisation through the effect of differential Faraday rotation. This effect is caused by the mixing of Faraday-rotating and synchrotron-emitting volumes, which leads to depolarisation of emission from the far end of the volume. This is why we expect to probe mostly local ISM at low radio frequencies.

Using the RM-synthesis\footnote{\url{https://github.com/sabourke/pyrmsynth_lite}} technique, we can separate polarised emission observed at different frequencies based on the amount of Faraday rotation along the LOS \citep{burn66, brentjens05}. For each LOS, RM-synthesis returns a Faraday spectrum, which describes how polarised intensity changes with Faraday depth.  The result of applying RM-synthesis to all LOS is the position-position-Faraday depth cube, or the Faraday cube, which represents the three-dimensional distribution of Faraday structures.

Faraday cubes of smaller fields of view, created using the low-resolution observations of polarised synchrotron emission by LOFAR, have already revealed a complex web of local magneto-ionic structures \citep{iacobelli13, jelic14, jelic15, vaneck17, turic21}. These discoveries were further advanced with the release of the LOFAR Two Metre Sky Survey \citep[LoTSS,][]{shimwell17, shimwell19, shimwell22}, covering a large part of the inner and outer Galaxy in the northern sky.  \citet{vaneck19} analysed the preliminary LoTSS data release in the HETDEX field. This was continued by \citet{erceg22} and \citet{erceg24} , who analysed the larger area in the high-latitude outer Galaxy using the LoTSS second data release (LoTSS-DR2). The Faraday cube mosaic created by \citet{erceg22} revealed structures in the local ISM in unprecedented detail. The authors found a high degree of order in the observed Faraday structures, primarily reflected in elongated depolarisation canals \citep[e.g.][]{sokoloff98, haverkorn04, jelic15}, which follow structures in polarised emission across the mosaic. \citet{erceg24} conducted a multi-tracer analysis on the LoTSS data and found alignment between the depolarisation canals, the plane-of-sky magnetic field and starlight polarisation. They used this alignment to estimate the distance to the structures observed by LoTSS to be around 200 pc. The research of \citet{turic21, erceg22} and \citet{erceg24} indicates that a large portion of the emission observed with LoTSS can be found within the first several hundreds parsecs.

Spectroscopic observations have revealed that the local ISM within the nearest 30 pc is described by the complex of local interstellar clouds (CLIC), which consist of numerous warm clouds surrounding the Sun \citep[e.g.][]{redfield08b, frisch11}. The closest one, the Local Interstellar Cloud \citep[LIC,][]{lallement92} was observed in spectra of most analysed stars and is thought to engulf the Solar system \citep{gry14}. Both the Sun and the CLIC are located inside the cavity of the Local Bubble, an irregular wall of cold and neutral gas carved in the ISM through multiple supernova explosions \citep{cox87}. The distance to the Local Bubble wall varies between 50 and 200 pc, depending on the direction of the observation. In some directions, the bubble opens and forms tunnels or chimneys to other cavities or the Galactic halo \citep{lallement19}. \citet{pelgrims20} created a data-based 3D model of the Local Bubble wall was created by using the dust extinction maps made by \citet{lallement19}. \citet{erceg24} used the distance predicted by this model to associate depolarisation canals observed by LoTSS with the Local Bubble. In this work, we use the model-predicted geometry of the Local Bubble wall to study the observed Faraday structures in the context of the Local Bubble and CLIC. 

This paper builds upon our previous work, this time focusing on a new large LoTSS mosaic, covering approximately 1400 square degrees in the high-latitude inner Galaxy. This paper is structured as follows. In Sect. \ref{sec:data}, we describe all the data used in the analysis and briefly describe the processing and mosaicing of the LoTSS-DR2 data. In Sect. \ref{sec:stat_tools}, we introduce the statistical tools used to analyse LoTSS data and quantify alignment between different data sets. In Sect. \ref{sec:faraday_cube}, we present and describe in detail the Faraday cube of LoTSS inner mosaic. Results of the multi-tracer analysis and estimates of the distance to the Faraday structures are presented in Sect. \ref{sec:multi_tracer}. In Sect. \ref{sec:discussion}, we associate the observed Faraday structures with the Local Bubble and analyse the implications of LoTSS data on the CLIC and the Bubble wall. We discuss the limitations of LoTSS observations in Sect. \ref{sec:limitations} and finally summarise our work and findings in Sect. \ref{sec:summary&conclusion}.

\section{Data and processing}\label{sec:data}

\begin{table} [!ht]
	\caption{Frequency range, corresponding resolution and maximum scale in Faraday space for a certain percentage of the fields used for the mosaic.}
	\begin{center}
		\begin{tabular}{lcccr}
			\hline 
			\hline
			Frequency & $\delta \phi$ &  $\Delta\phi_{\mathrm{scale}}$ & Field & Color in  \\
			 range  &  &   & percentage  & Fig. \ref{coverage} \\
			$[$MHz] & [$\mathrm{rad~m^{-2}}$] & [$\mathrm{rad~m^{-2}}$] & [\%] &  \\
			\hline 
			120$-$167 & 1.16 & 0.97 & 96.2 & gray \\  
			120$-$165 & 1.20 & 0.95 & $\ \ $2.7 & yellow \\  
			120$-$163 & 1.23 & 0.93 & $\ \ $1.1 & orange \\ 
			\hline
		\end{tabular} 
	\end{center}
    \tablefoot{ The values change from field to field due to different frequency ranges. Different fields are indicated with coloured circles in Fig. \ref{coverage}. Here we include only the 186 fields used to make the final mosaic.}
	\label{polja}
\end{table}

\begin{figure}[!ht]
  \centering
    \includegraphics[width=\hsize]{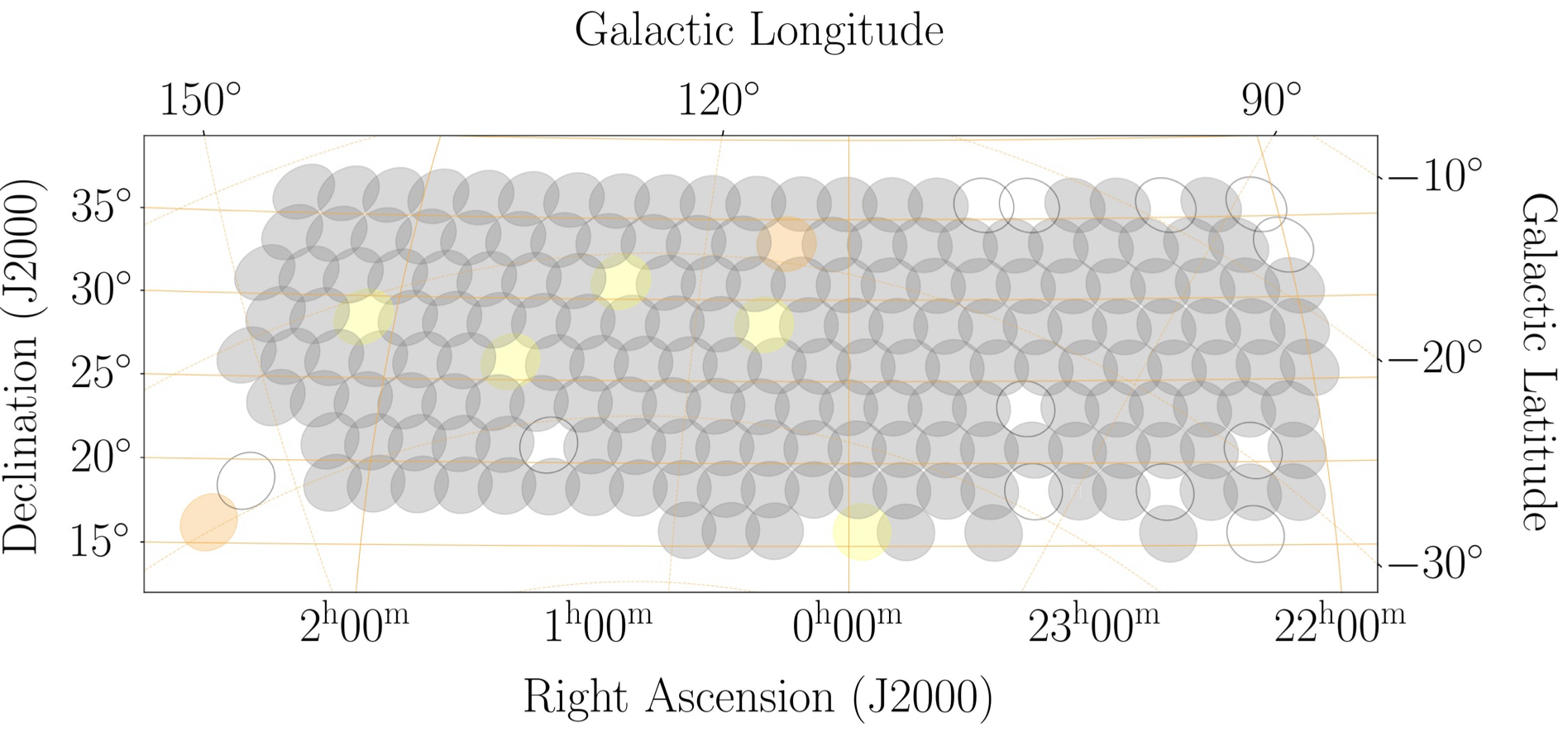}
  \caption{Mosaic coverage. Circles represent the primary beam size of each observation. Grey, yellow and orange circles denote observations in the frequency ranges of 120 $-$ 167 MHz, 120 $-$ 165 MHz and 120 $-$ 163 MHz, respectively. The empty circles represent the pointings which were discarded due to high noise.}
  \label{coverage}
\end{figure}

\begin{figure}
   \centering
   \includegraphics[width=\hsize]{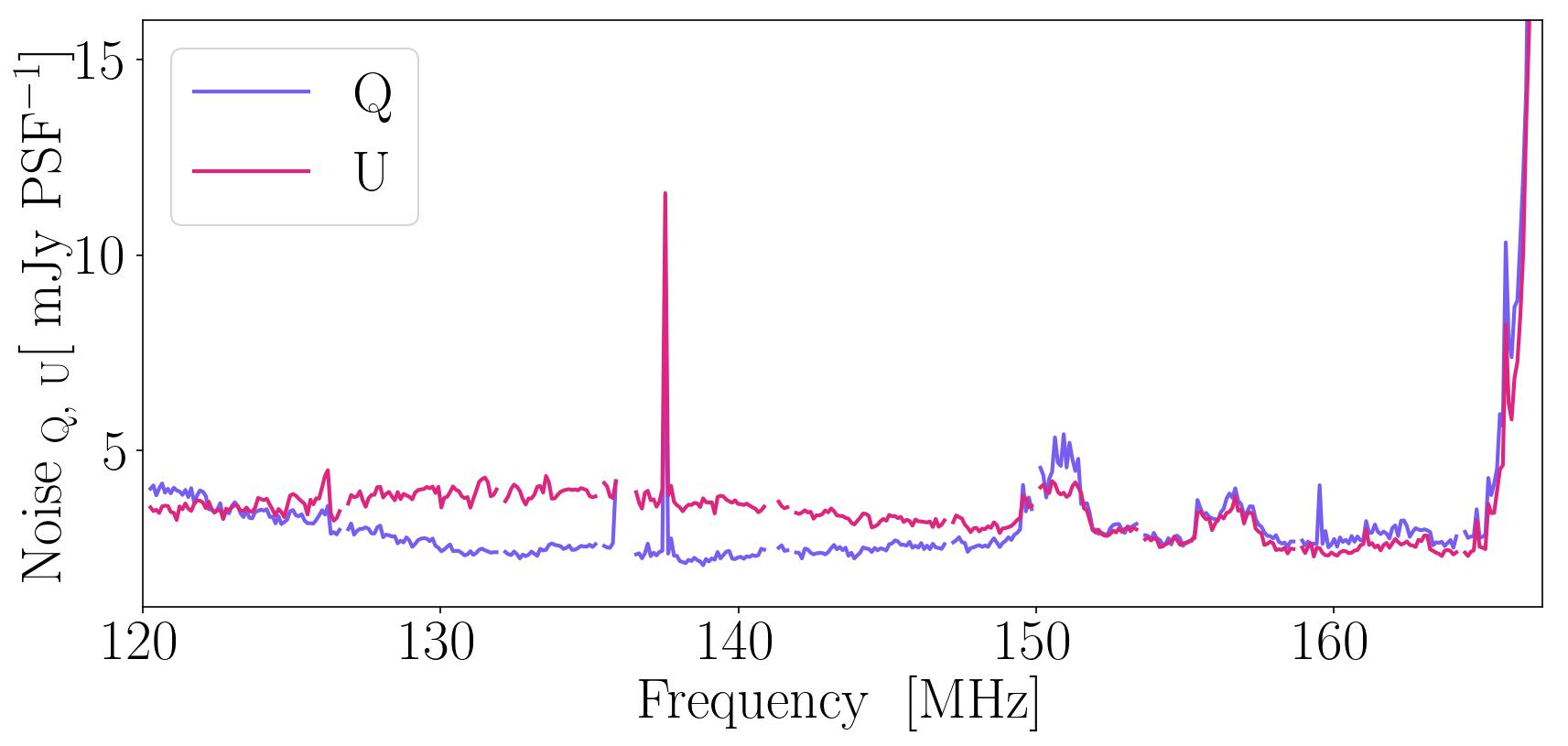}
      \caption{Noise in Stokes $Q$ and $U$ for a randomly chosen frequency cube that was not corrected for the primary beam.}
         \label{rnd_noise}
\end{figure}

\begin{figure}
   \centering
   \includegraphics[width=\hsize]{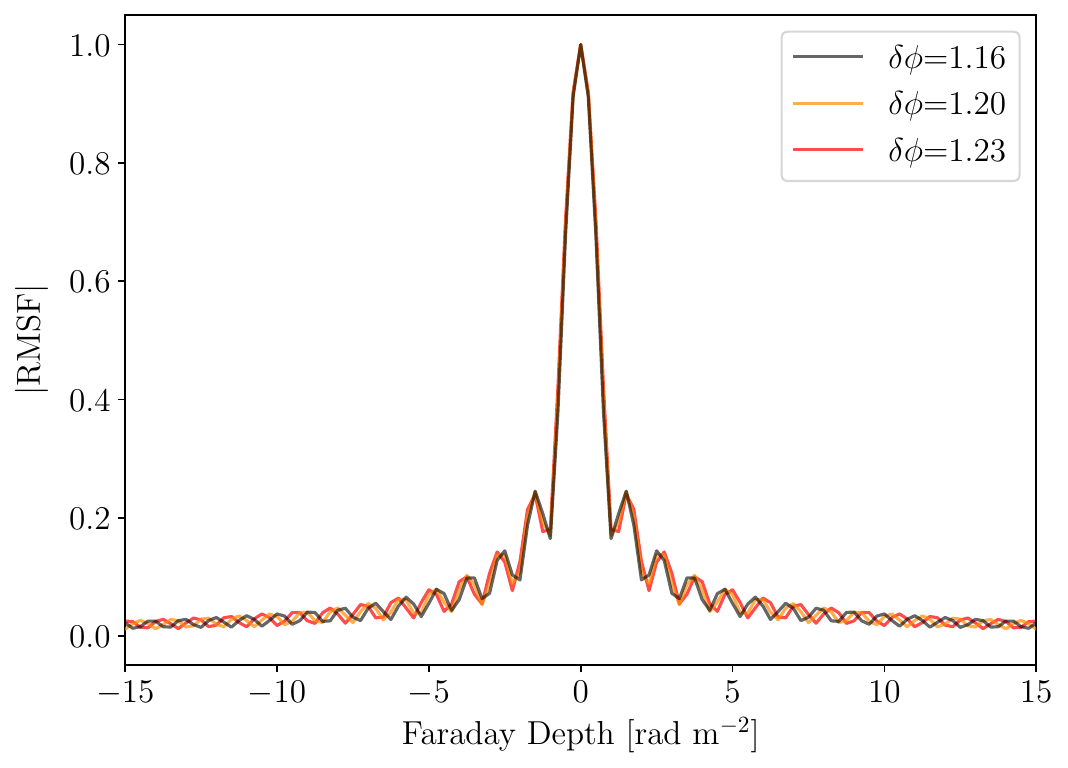}
      \caption{RMSF functions of observations with different frequency ranges (see Table \ref{polja}). The difference between the functions is marginal. 
              }
         \label{rmsf}
\end{figure}

\begin{figure}
   \centering
   \includegraphics[width=\hsize]{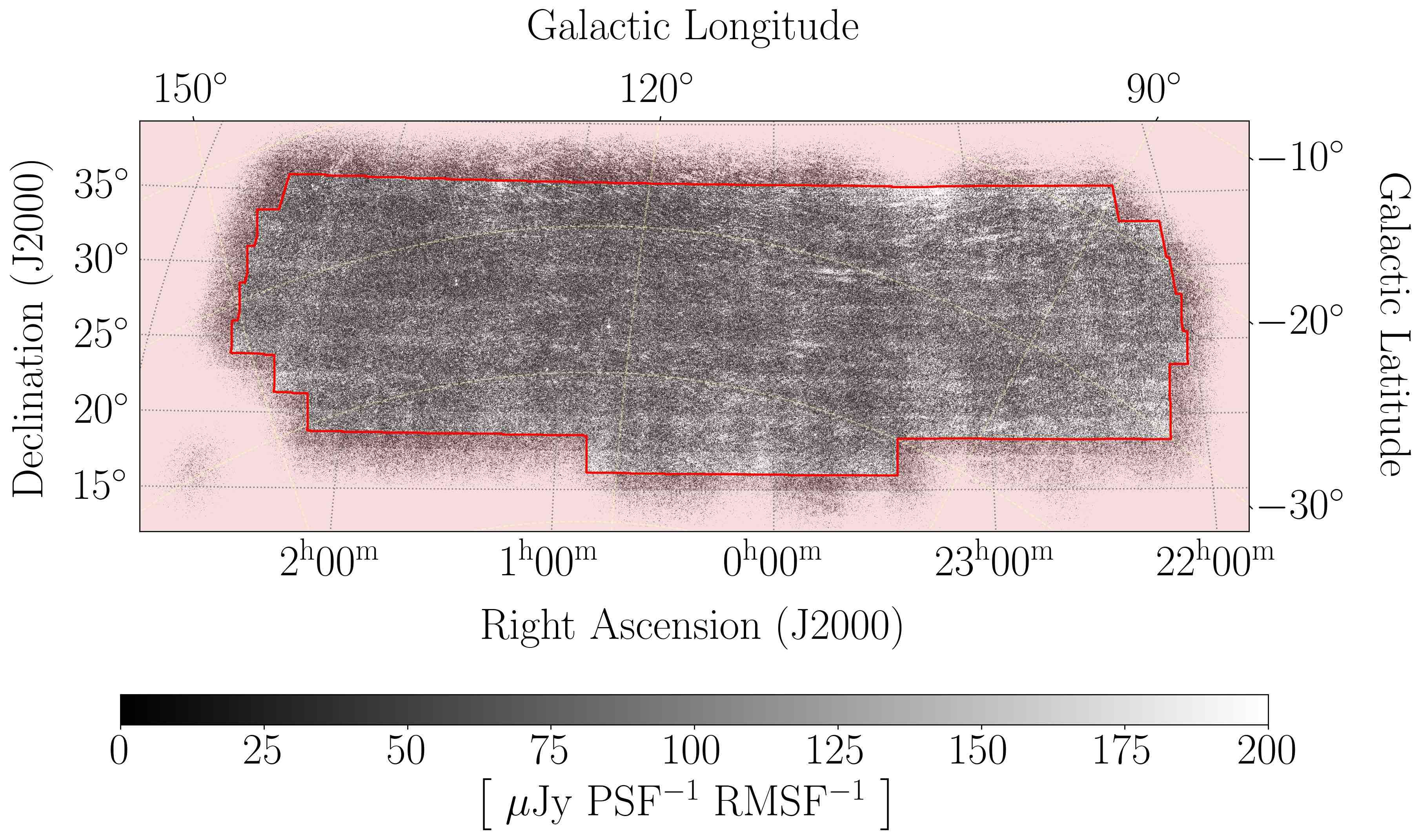}
      \caption{Mosaic at a noise dominated Faraday depth, $\phi = 50 \ \mathrm{rad \ m^{-2}}$. The red shaded area represents the mask used in the calculation of Faraday moments in Sect. \ref{sec:moments}.}
         \label{mask}
\end{figure}

\begin{figure}
   \centering
   \includegraphics[width=\hsize]{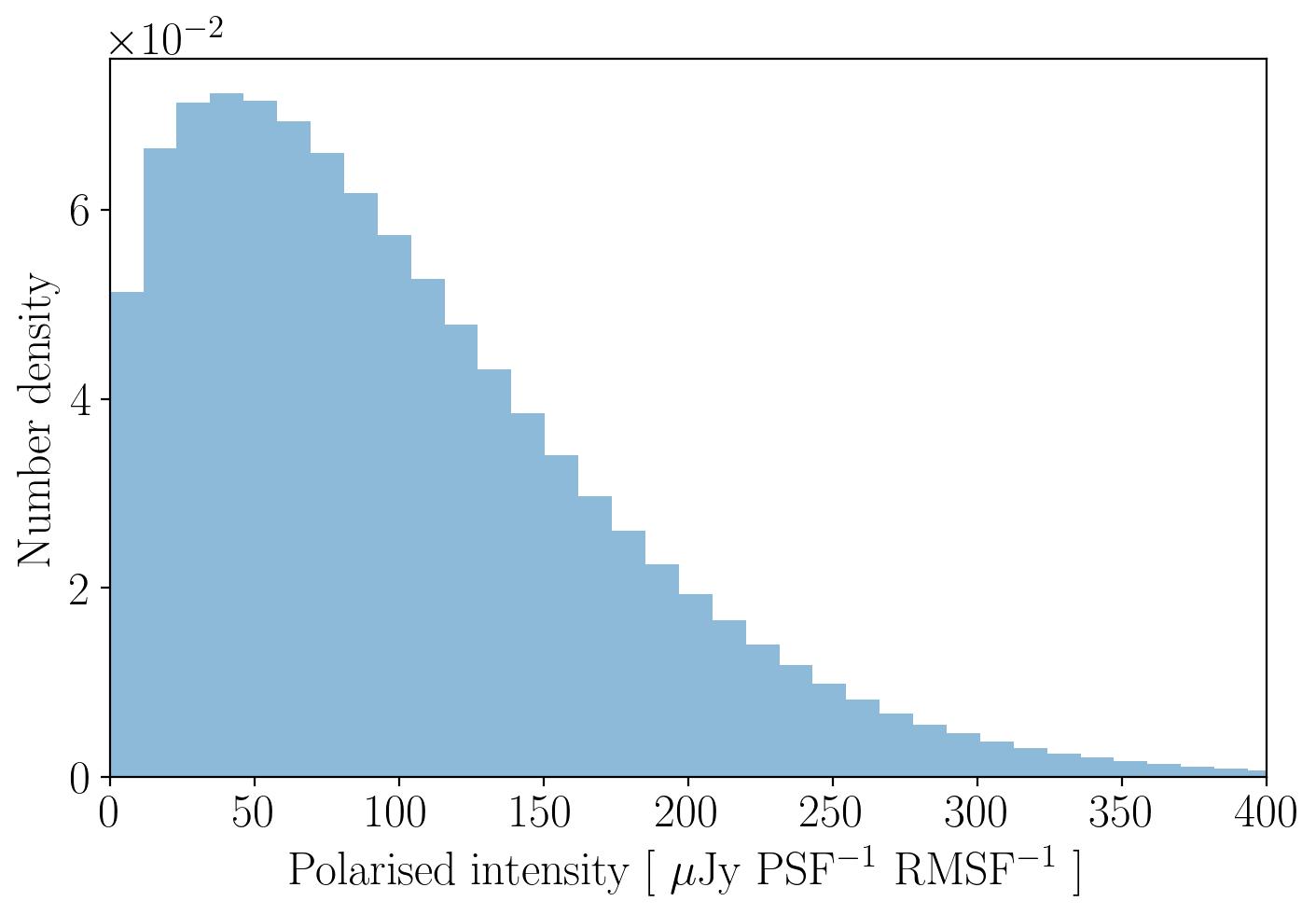}
      \caption{Histogram of polarised intensity at Faraday depth $\phi = 50 \ \mathrm{rad \ m^{-2}}$. The noise level in the mosaic can be estimated from the width of this distribution as 110 $\rm{\mu Jy \ PSF^{-1} \ RMSF^{-1}}$, while the mean of the distribution represents the polarisation bias, 150 $\rm{\mu Jy  \ PSF^{-1} \ RMSF^{-1}}$.}   \label{noise_hist}
\end{figure}

In this section, we present the LoTSS-DR2 data used in this work and briefly outline the processing and mosaicing procedures. The data, along with all the products described in Sect. \ref{sec:faraday_cube} are publicly available\footnote{The mosaicked Faraday cube and other derived products of this paper (maximum polarised intensity image described in Sect. \ref{subsec:mosaic cube} and Faraday moments described in Sect. \ref{sec:moments}) are publicly available at FULIR database \url{https://data.fulir.irb.hr/islandora/object/irb:405}.}. Additionally, we describe other data sets used in our multi-tracer analysis.

\subsection{LoTSS very low-resolution images}

LoTSS-DR2 \citep{shimwell22} covers $5634$ square degrees of the Northern sky with observations distributed in two large regions, centred at $RA=12^{\rm{h}}$ and $RA=0^{\rm{h}}$. LoTSS-DR2 provides images in full Stokes parameters at different resolutions. 

In this work, we focus on a large region centred at $RA=0^{\rm{h}}$. This region consists of 198 pointings, covering approximately $1400$ square meters of the sky (Fig. \ref{coverage}), which in Sect. \ref{sec:mosaicing} we combine into a mosaic. Each pointing was observed for $\sim$8h during different LOFAR observing cycles. More information on these observations and their calibration can be found in \citet{shimwell22}. We use primary beam corrected and uncorrected Stokes $Q$ and $U$ very-low (vlow) resolution images, which are observed using the LOFAR High Band Antennas \citep[HBA,][]{vanhaarlem13} in the frequency range from 120 to 167 MHz, with a step of 0.098 MHz. The observed frequency range changes slightly over the pointings, which has a minimal impact on the final mosaic (see Table \ref{polja} and Fig. \ref{coverage}). The resolution of images varies between 4 and 5.5 arcmin; however, to produce the mosaic we use images at a common resolution of 5.5 arcmin. Throughout the paper, we use the term 'LoTSS' to refer to LoTSS vlow polarisation data.

An example of noise as a function of frequency in primary beam uncorrected Stokes $Q$ and $U$ images for a randomly chosen pointing is shown in Fig. \ref{rnd_noise}. The level of noise in Stokes $Q$ and $U$ is comparable. Broad radio-frequency interferences (RFIs) at 150-160 MHz are shared among multiple fields \citep{offringa13}, as is the RFI seen around 140 MHz. We flag all the channels in which noise exceeds the mean value of noise over the full bandwidth of the given observation by 4 sigma or more. Not all flagged and missing channels are shared between all observations; however, this does not have a strong effect on the produced Faraday cubes.

\subsection{RM-synthesis}

We applied RM-synthesis on each field separately before combining them into a mosaic. The resulting Faraday cubes span a Faraday depth range from $-50$ to $+50~\mathrm{rad~m^{-2}}$, with a step of $0.25~\mathrm{rad~m^{-2}}$. 

The resolution in Faraday space, $\delta\phi$, is defined as the width of the RM spread function (RMSF, Fig. \ref{rmsf}) and is constrained by the wavelength range $\Delta\lambda$ of observations as $\delta\phi\approx 2\sqrt{3}/\Delta\lambda^2$. As the wavelength range varies between observations, the resolution in Faraday depth also varies slightly (see Figs. \ref{coverage} and \ref{rmsf}). However, the majority (96.2\%) of observations share a common resolution of $\delta \phi\approx~1.15~\mathrm{rad~m^{-2}}$, with negligible differences in the remaining fields (Table \ref{polja}). 

Another relevant constraint of RM-synthesis is the shortest observed wavelength $\lambda_\mathrm{min}$, which limits the largest resolvable Faraday structure to $\Delta\phi_\mathrm{scale}\approx\pi/\lambda^2_\mathrm{min}\approx~0.97~\mathrm{rad~m^{-2}}$. In LoTSS observations, the size of the largest resolvable structure is comparable to the resolution. Therefore, we can only detect narrow, Faraday thin structures or the edges of wide, Faraday thick structures in LoTSS Faraday spectra, as described in \citet{brentjens05} and \citet{vaneck17}.

The noise in the Faraday cubes follows a Rician distribution \citep[e.g.][]{brentjens05, hales12} and can be estimated by multiplying the standard deviation of the polarised intensity at any noise-dominated Faraday depth by $\sqrt{2}$. We make this calculation for each Faraday cube at $50~\mathrm{rad~m^{-2}}$ and remove fields in which the noise is higher than 150 $\mathrm{\mu Jy~PSF^{-1}~RMSF^{-1}}$ (empty grey circles in Fig. \ref{coverage}). Finally, we are left with 186 fields to connect into a mosaic.

\subsection{Mosaicing}
\label{sec:mosaicing}
We used \texttt{MontagePy}\footnote{\url{http://montage.ipac.caltech.edu/docs/montagePy-UG.html}}, a python expansion of the Astronomical Image Mosaic Engine \texttt{Montage}\footnote{\url{http://montage.ipac.caltech.edu}}, and followed the mosaicing procedure outlined in \citet{erceg22}. 
\texttt{Montage} provides the tools to combine multiple images into a mosaic on a curved grid by conserving flux while changing pixel areas during the reprojection. This is achieved by assigning a weight to each input pixel based on the amount of surface overlap with the output pixel. In addition to this, we used a weighing scheme described in \citet{erceg22}. We assigned weight to each field according to its noise level and we assigned weight to each pixel, according to its position with regards to the primary beam of the observation. We first created separate mosaics from Stokes $Q$ and $U$ images and then calculated the final mosaic Faraday cube as $P=\sqrt{Q^2+U^2}$. 

The noise in the final polarised intensity mosaic is not uniform but does not vary much (Fig. \ref{mask}). We estimated the noise from the distribution of intensity at $50~\mathrm{rad~m^{-2}}$ (Fig. \ref{noise_hist}), as described in the previous section. The noise level is $110~\mathrm{\mu Jy~PSF^{-1}~RMSF^{-1}}$ and the polarisation bias is $150~\mathrm{\mu Jy~PSF^{-1}~RMSF^{-1}}$. We did not correct for this bias, as its impact on the quantitative analysis of this paper is negligible.

\subsection{Other data sets}
In the paper, we investigate alignment between the orientation of depolarisation canals in the mosaic and other ISM tracers - \textit{Planck} magnetic field, HI filaments and the starlight polarisation, following the same analysis as in \citet{erceg24}. All the data sets used are constrained to the area covered by the inner mosaic.

\textit{Magnetic field.} We derived the plane-of-sky magnetic field orientation from the Planck 353 GHz Stokes $Q$ and $U$ polarization maps \citep{planck15, planck16a}, which are publicly available at the Planck Legacy Archive\footnote{\url{https://pla.esac.esa.int}}. This data set traces the polarisation angle of dust emission, which is rotated by $90^\circ$ to the plane-of-sky magnetic field. To compensate for the lower signal-to-noise ratio of the observed polarised dust emission, we smoothed the \textit{Planck} data to a resolution of 1 degree.

\textit{Starlight polarisation.} We used a compilation of optical polarisation catalogues published by \citet{panopoulou23}. This catalogue contains data from 81 separate publications with measurements of starlight polarisation and distances to the observed stars, obtained by cross-matching with the \citet{bailerjones21} catalogue based on the Gaia Early Data Release 3. We discarded sources with starlight polarisation errors larger than 60 degrees, which left us with 117 sources in our field. 

\textit{HI brightness temperature.} HI spectroscopic data in this paper come from the Effelsberg-Bonn HI survey\footnote{\url{http://cdsarc.u-strasbg.fr/viz-bin/qcat?J/A+A/585/A41}} \citep[EBHIS, ][]{winkel16}. The position-position-velocity (PPV) cubes cover a local-standard-of-rest velocity range of $|v_\mathrm{LSR}| < 600~\mathrm{km~s^{-1}}$ with a spectral resolution of $1.44~\mathrm{km~s^{-1}}$ and an angular resolution of 10.8 arcmin. We found significant emission in velocity range from -13.9 to 120~$\mathrm{km~s^{-1}}$, mostly organised in HI filaments. We expect that the structures observed by LoTSS are mostly at a distance of several hundred parsecs. In an effort to include only the local HI gas, we avoid intermediate to high velocity cold neutral clouds that are expected to occupy distant halo, at distances greater than 1 kpc \citep[e.g.][and references therein]{marasco22}. For this reason, we focused on HI emission integrated over the velocity range of $|v_\mathrm{LSR}| < 30 \mathrm{km~s^{-1}}$. 

\section{Statistical tools}
\label{sec:stat_tools}
In this section, we describe the main statistical tools used in the analysis. As in \citet{erceg24}, the Rolling Hough transform is used to quantify the orientation of depolarisation canals and HI filaments. The alignment between different tracers is analysed with Projected Rayleigh statistics. Additionally, we outline the calculation of statistical Faraday moments of LoTSS data.

\subsection{Rolling Hough transform}

Rolling Hough transform (RHT) \citep{clark14} is a computer vision algorithm that can be used to detect and quantify linear features in images of the ISM, such as depolarisation canals or HI filaments. It is based on the Hough algorithm \citep{hough62} and works on a pixel-per-pixel basis, by calculating the probability that each pixel is a part of a coherent linear structure, $R(\theta, x, y)$. The RHT algorithm requires three parameters to define the properties of lines we want to detect. The probability threshold (Z) sets the minimum acceptable probability that a pixel is part of a linear structure, the window diameter ($D_w$) determines the length of the shortest detectable line. The smoothing kernel diameter ($D_K$) defines the size of a two-dimensional kernel that is convolved with the original image to smooth out large scales and sharpen linear features through unsharp masking. Following the approach by \citet{clark14} we obtain a distribution of the orientation of structures in a specific area and integrate the RHT distributions over it:

\begin{equation}
    \Tilde{R}(\theta) = \int \int_{area} R(\theta, x, y) \ dxdy.
\end{equation}

To calculate the mean $\langle \theta \rangle$ and the distribution spread, $\delta\theta$, we follow \citet{jelic18}. This is done by projecting the angle $\theta$ to a full circle and defining every point as a vector of length $\Tilde{R}^2(\theta) d\theta$. The authors define a vector S:

\begin{equation}
    S = \frac{\int_{-\pi/2}^{\pi/2} \Tilde{R}^2(\theta)e^{2i\theta}d\theta}{\int_{-\pi/2}^{\pi/2} \Tilde{R}^2(\theta)d\theta}, 
\end{equation}
whose direction and length can be used to measure $\langle \theta \rangle$ and $\delta\theta$:

\begin{equation}
     \langle \theta \rangle = \frac{1}{2}\rm{arctan}\frac{\rm{Im}(S)}{\rm{Re}(S)},
\end{equation}

\begin{equation}
     \delta\theta = \frac{1}{2}\sqrt{\rm{ln}(1/|S|^2)}.
\end{equation}
The mean value of the distribution is not strongly affected by the choice of parameters; however, it affects the standard deviation by making the final result more or less noisy, as discussed in \citet{jelic18}.

\subsection{Projected Rayleigh statistics}

We quantified the alignment between two tracers by using the Projected Rayleigh statistics \citep[PRS,][]{jow18}. This statistic evaluates whether two tracers are aligned by testing whether the distribution of their orientation angle differences peaks around zero. In this paper, as in \citet{erceg24}, we adopted the approach outlined by \citet{panopoulou21} to compute the PRS and estimate the uncertainty and the statistical significance of the alignment. PRS is defined as:

\begin{equation}
    PRS = \frac{1}{\sqrt{\sum_i^N\frac{w_i^2}{2}}}\sum_i^Nw_i \cos2\Delta\psi_i,
\end{equation}
where the angle difference $\Delta\psi$ between two angles $\theta_1$ and $\theta_2$ is given by:
\begin{equation}\label{eq:deltapsi}
    \Delta \psi = \frac{1}{2} \arctan \left( \frac{\sin2\theta_1 \cos2\theta_2-\cos2\theta_1 \sin 2\theta_2}{\cos2\theta_1\cos2\theta_2+\sin2\theta_1\sin2\theta_2}\right),
\end{equation}
to account for $\pi$-ambiguity \citep{planck16c}. We defined weights $w_i$ as
\begin{equation}
    w_i = \frac{1}{\sigma_{\Delta \psi}^2},
\end{equation}
where $\sigma_{\Delta \psi}$ is estimated using a Monte Carlo (MC) simulation and includes uncertainties in both angle measurements, $\sigma_{\theta_1}$ and $\sigma_{\theta_2}$. A positive PRS value indicates alignment, while a negative one indicates that the two tracers are perpendicular.

The uncertainty in the PRS is estimated by drawing a sample of angles from normal distributions with $\mu = \Delta \psi$ and $\sigma = \sigma_{\Delta \psi}$ and repeating this 10,000 times. We calculate the PRS for each iteration and create a distribution, whose standard deviation we report as the PRS uncertainty. To evaluate the statistical significance of the alignment, we used another MC method as outlined in \citet{panopoulou21}. We simulate samples of random angles observed with the same uncertainty as the real data. More precisely, from a uniform distribution of angles in range $[-90^\circ, +90^\circ]$ we randomly draw angles that represent mean values of normal distributions with standard deviations equal to $\sigma_{\Delta \psi}$ values. We repeat this procedure 10,000 times and in each iteration calculate the PRS value. The 99.9th percentile of the obtained PRS distribution represents the significance threshold. A PRS value surpassing this threshold indicates statistical significance at a level of $3\sigma$ or higher.

\subsection{Faraday moments}
3D data cubes can be visualised in a compact way by using statistical moments. Through moments of Faraday spectra, we can more accurately characterise complex spectra containing multiple peaks. The zeroth Faraday moment, $M_0$, represents the total integrated intensity of polarised emission over the whole Faraday spectrum. The first moment, $M_1$, is the average Faraday depth weighted by the polarised intensity and the second moment, $M_2$, represents the variance of the Faraday spectrum weighted by the polarised intensity. To compute the moments from the LoTSS mosaic, we use the definitions of the Faraday moments from \citet{dickey19}:

\begin{equation}
    \begin{aligned}
    M_0 & = \sum_{i=1}^n T_i \ d\phi, \\
    M_1 & = \frac{\sum_{i=1}^n T_i \cdot \phi_i}{\sum_{i=1}^n T_i}, \\
    M_2 & = \frac{\sum_{i=1}^n T_i \cdot (\phi_i - M_1)^2}{\sum_{i=1}^n T_i}.
    \label{moments}
  \end{aligned}
\end{equation}

Statistical moments, particularly second and higher moments, are highly sensitive to noise contamination. In \citet{erceg22} we computed moments using a constant noise threshold at 5 $\sigma$ above the mean noise level. Here, we improve on that approach by using an adaptive noise threshold. For each Faraday spectrum, we calculate the mean level of noise and set the threshold to 4 $\sigma$ above it. This approach ensures that we are not cutting low-intensity emission by setting the threshold too high. As in \citet{erceg22}, we masked the mosaic's edges, where the noise is boosted by the primary beam (Fig. \ref{mask}). We defined the mask by discarding the area where there are no overlaps between individual pointings.

\section{Faraday cube}
\label{sec:faraday_cube}

\begin{figure*}
   \centering
   \includegraphics[width=0.97\textwidth]{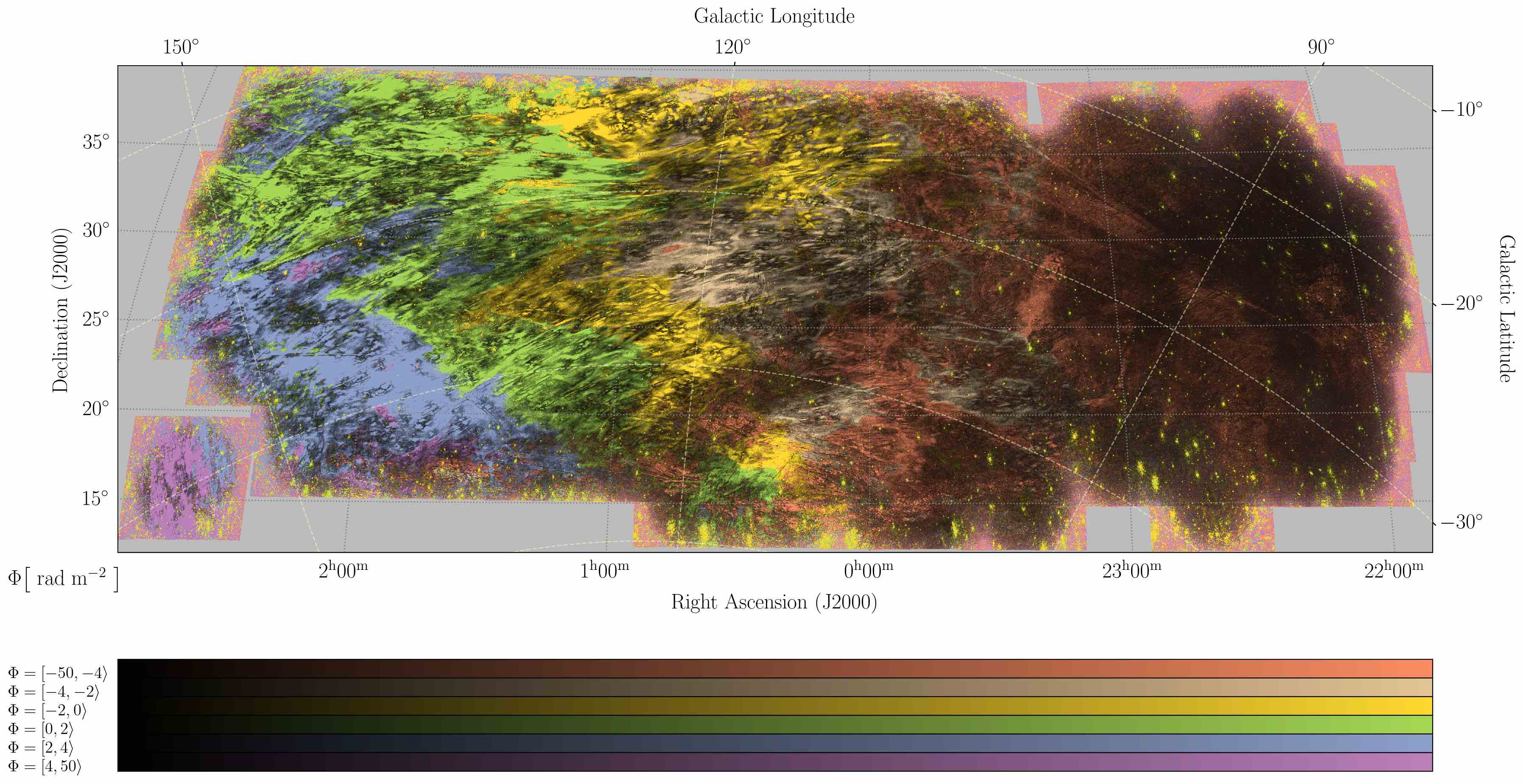}
      \caption{Image of maximum polarised intensity of LoTSS-DR2 mosaic, with different colours representing different ranges of Faraday depths, within which peaks are located. }
         \label{mosaic}
\end{figure*}

In this section, we present the mosaic Faraday cube of polarised synchrotron emission in the inner Galaxy. The resulting Faraday cube spans an area from $21^{\rm{h}}45^{\rm{m}}$ to $3^{\rm{h}}$ in $RA$, from $\mathrm{12^\circ}$ to $\mathrm{40^\circ}$ in $Dec$, and from $-50$ to $+50~\mathrm{rad~m^{-2}}$ in Faraday depth. Chosen slices from the Faraday cube are presented in Appendix \ref{app:slices}. Additionally, we present a 2D representation of the cube by utilising the Faraday moments.

\subsection{The LoTSS mosaic Faraday cube} \label{subsec:mosaic cube}

We observe synchrotron emission across a Faraday depth range spanning from approximately -50 to +8 $\mathrm{rad~m^{-2}}$. A small amount of patchy, disconnected emission with typical values lower than 1 $\mathrm{mJy~PSF^{-1}~RMSF^{-1}}$ is present throughout the mosaic at Faraday depths from $-50~\mathrm{rad~m^{-2}}$ to $-15~\mathrm{rad~m^{-2}}$. As Faraday depth increases from  $-15~\mathrm{rad~m^{-2}}$ towards $+8~\mathrm{rad~m^{-2}}$, the synchrotron emission organises in multiple fronts that 'travel' from roughly $23^\mathrm{h}00^\mathrm{m}$ towards east. This creates a large gradient over roughly two-thirds of the mosaic area. The number, thickness, morphology and velocity of the fronts change as the gradient evolves. The fronts travel more slowly at Faraday depths below $-5~\mathrm{rad~m^{-2}}$, on average around $1^\circ$ eastward per $\mathrm{rad~m^{-2}}$. From -5 to +8 $\mathrm{rad~m^{-2}}$ the fronts travel faster, around $4^\circ$ towards east per $\mathrm{rad~m^{-2}}$. The changing sign of the Faraday depth in the gradient likely indicates a curved morphology of the magnetic field. The parallel component of the magnetic field in the west part of the mosaic is oriented away from the observer; however towards the east the orientation changes and the parallel component points towards the observer. The gradient in Faraday depths is accompanied by a gradient in emission, with intensity increasing from west to east (see Appendix \ref{app:slices}). Around Faraday depth of $0~\mathrm{rad~m^{-2}}$, we can see strong emission from several instrumentally polarized point sources.

In Fig. \ref{mosaic}, we represent the 3D Faraday cube as a 2D maximum polarised intensity image by choosing only the highest peak along each LOS. Different colours represent different Faraday depths at which the peak is found. This image highlights the extent and the morphology of the gradient described above. Mean maximum intensity in the mosaic is $\mathrm{2~mJy~PSF^{-1}~RMSF^{-1}}$. The emission is organised in filamentary structures oriented in a roughly east-west direction and often accompanied by straight and narrow depolarisation canals following the same orientation. On top of this emission, there is a less ordered emission, characterised by curvy depolarisation canals, which reveals a more complex and turbulent nature of ISM on smaller scales. The emission is notably weaker in the western part of the mosaic ($RA\leq\mathrm{23^h15^m}$), where the average maximum polarised intensity is $\mathrm{0.7~mJy~PSF^{-1}~RMSF^{-1}}$. This could be caused by weaker synchrotron emission in that direction or by stronger depolarisation by foreground than in the rest of the mosaic. 

A single isolated field in the bottom east part of the mosaic shows a somewhat different morphology than the rest of the mosaic. Structure in this field is ordered and Faraday depths at which the emission appears are consistent with the gradient in Faraday depths; however, the main orientation of structures is rotated from the one in the rest of the mosaic by roughly 90 degrees.

\subsection{Faraday moments}\label{sec:moments}

\begin{figure*}
\centering
\includegraphics[width=0.9\textwidth]{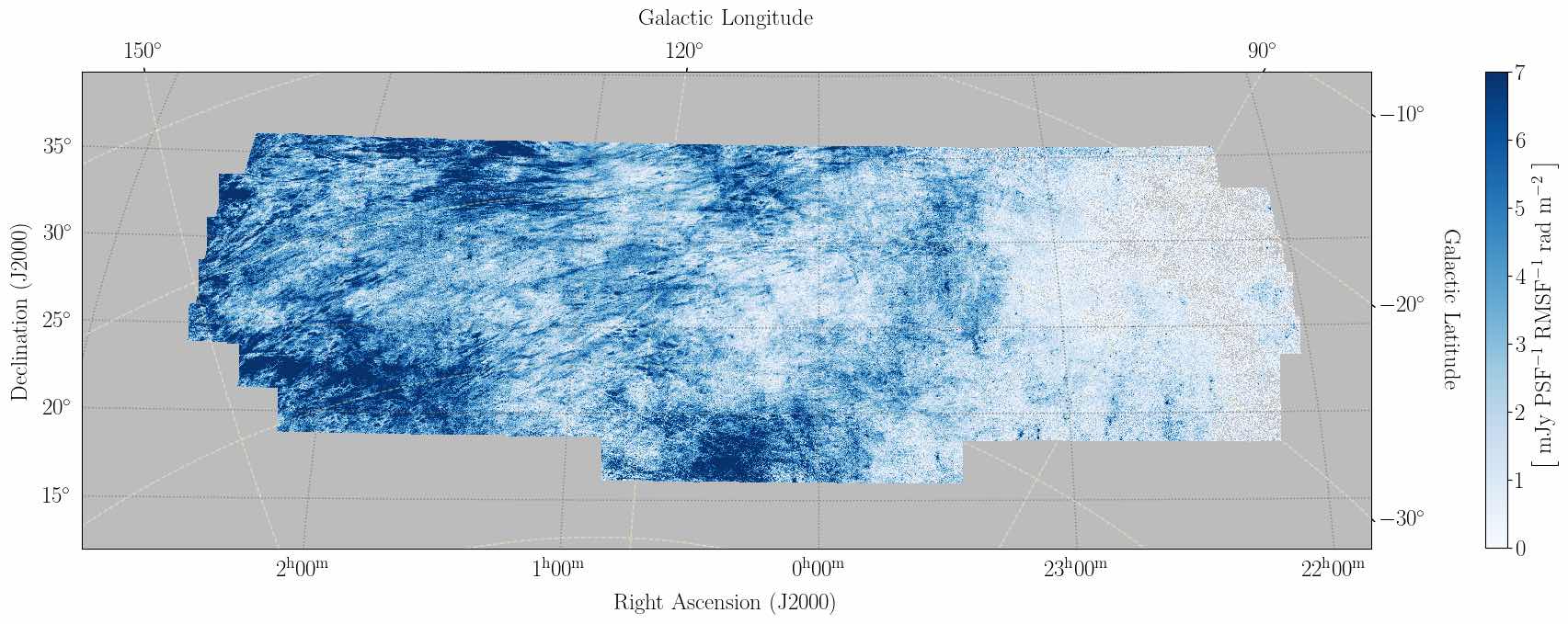}
\includegraphics[width=0.9\textwidth]{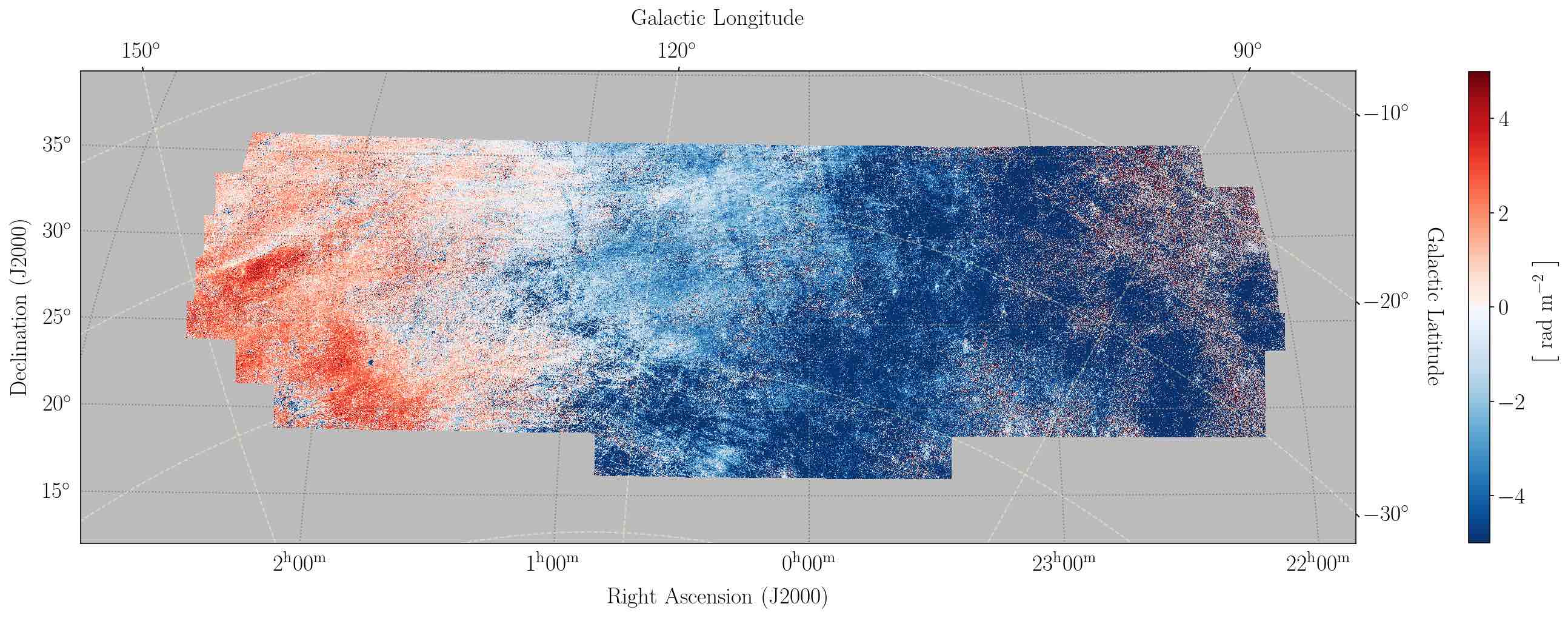}
\includegraphics[width=0.9\textwidth]{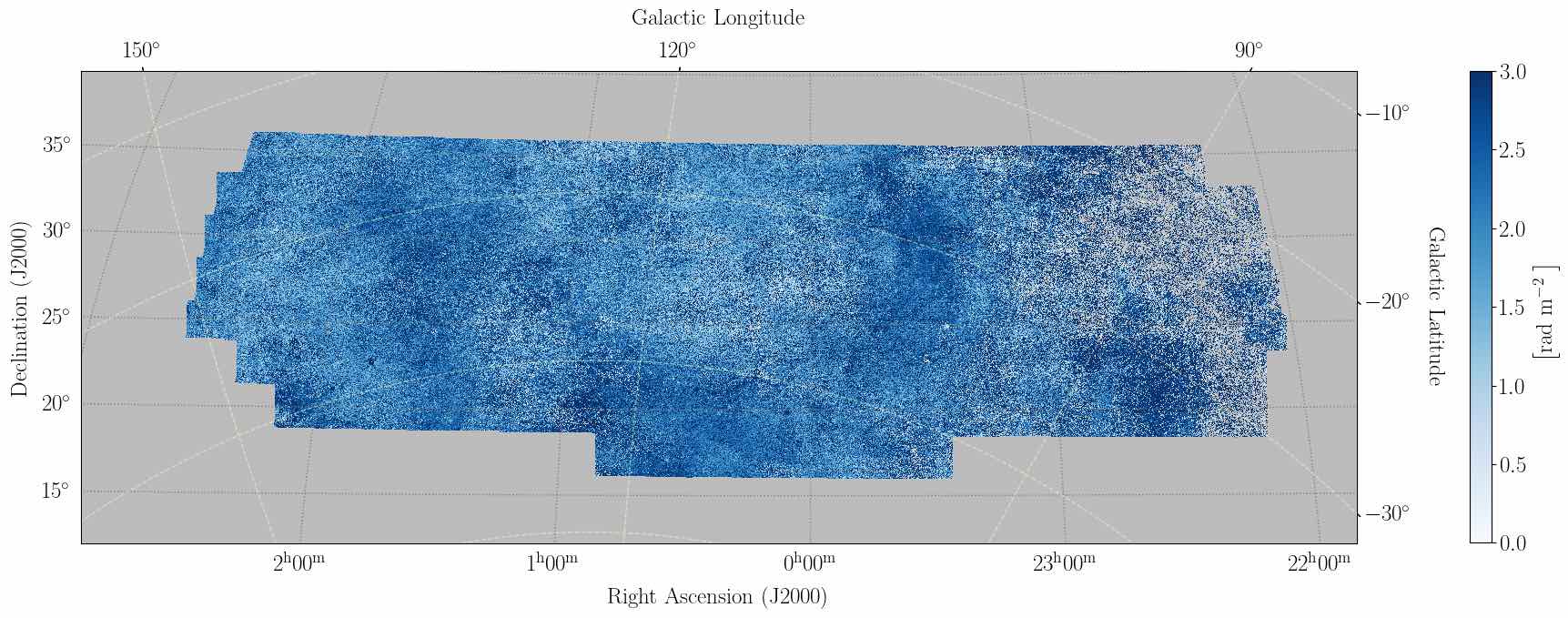}
\vspace{-0.2cm}
\caption{Maps of zeroth $M_0$, first $M_1$ and the square root of second $m_2$ Faraday moment (top to bottom, respectively). Black circles mark the location of Faraday spectra in Fig. \ref{FarSpec+m2}.}\label{M_slika}
\end{figure*}

\begin{figure}
   \centering
   \includegraphics[width=\hsize]{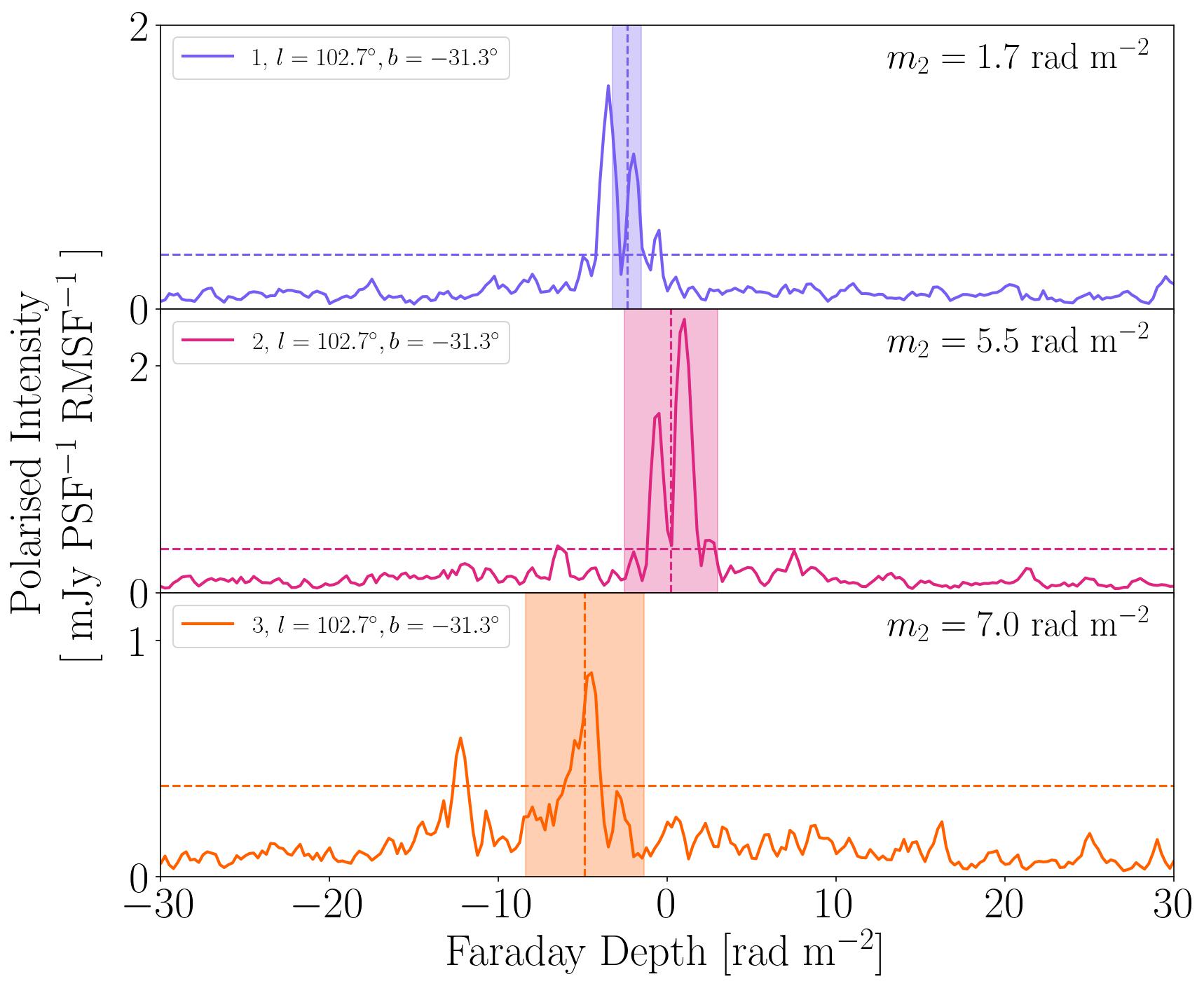}
      \caption{Example Faraday spectra. Vertical dashed line marks M1 and the width of the shaded area corresponds to m2. Horizontal dashed line marks the threshold used to calculate the moments. The locations from which the spectra were extracted are marked in Fig. \ref{M_slika}.} 
         \label{FarSpec+m2}
\end{figure}

The maps of zeroth, first and the square root of the second moment, $\rm{m_2}$, are presented in Fig. \ref{M_slika}. Zeroth moment (Fig. \ref{M_slika}, top panel) resembles the maximum polarised intensity image (Fig. \ref{mosaic}), which indicates that most of the Faraday spectra in the mosaic are dominated by a single peak. However, the width of the spectra, which is represented by the $\rm{m_2}$ map (Fig. \ref{M_slika}, bottom panel), is on average around $6~\mathrm{rad~m^{-2}}$, which is greater than width of a single peak. This points to a presence of lower-intensity secondary peaks, which accompany the dominating peak but do not significantly contribute to integrated total intensity of M0. The varying difference in Faraday depth between these peaks, along with the choice of thresholds, creates some of the variations in the $\rm{m_2}$ map. An example of this behaviour can be seen in Faraday spectra in Fig. \ref{FarSpec+m2}.

In the mostly uniform $\rm{m_2}$ map, the most notable exception is an arc of enhanced values between $0^{\rm{h}}30^{\rm{m}}$ and $23^{\rm{h}}30^{\rm{m}}$. This arc is faint in the maximum polarised intensity image (Fig. \ref{mosaic}) and more prominent in the M1 image, which points towards Faraday spectra with multiple peaks of comparable intensity. We found the emission connected to the arc in the Faraday cube between $-12$ and $-10~\mathrm{rad~m^{-2}}$, whose mean intensity is around $\mathrm{1~mJy~PSF^{-1}~RMSF^{-1}}$. The presence of this secondary structure with comparable intensity is in line with a jump towards negative M1 values in this area, that stands out from the main gradient (see middle panel of Fig. \ref{M_slika}). 

The eastern part of the mosaic is dominated by emission following the gradient in Faraday depths. The M1 map clearly shows the gradient changing sign over an irregular border, between $RA$ of $1^{\rm{h}}$ and $1^{\rm{h}}20^{\rm{m}}$. On top of this gradient, there are several narrow filaments and patchy structures with negative values in the M1 map, which correspond to weak and disconnected structures seen in the cube between $-20$ and $-5~\mathrm{rad~m^{-2}}$. These structures do not have a high signal-to-noise ratio, which limits their impact on the other two moment maps. 

\section{Multi-tracer analysis}
\label{sec:multi_tracer}
\begin{figure*}
   \centering
   \includegraphics[width=0.9\textwidth]{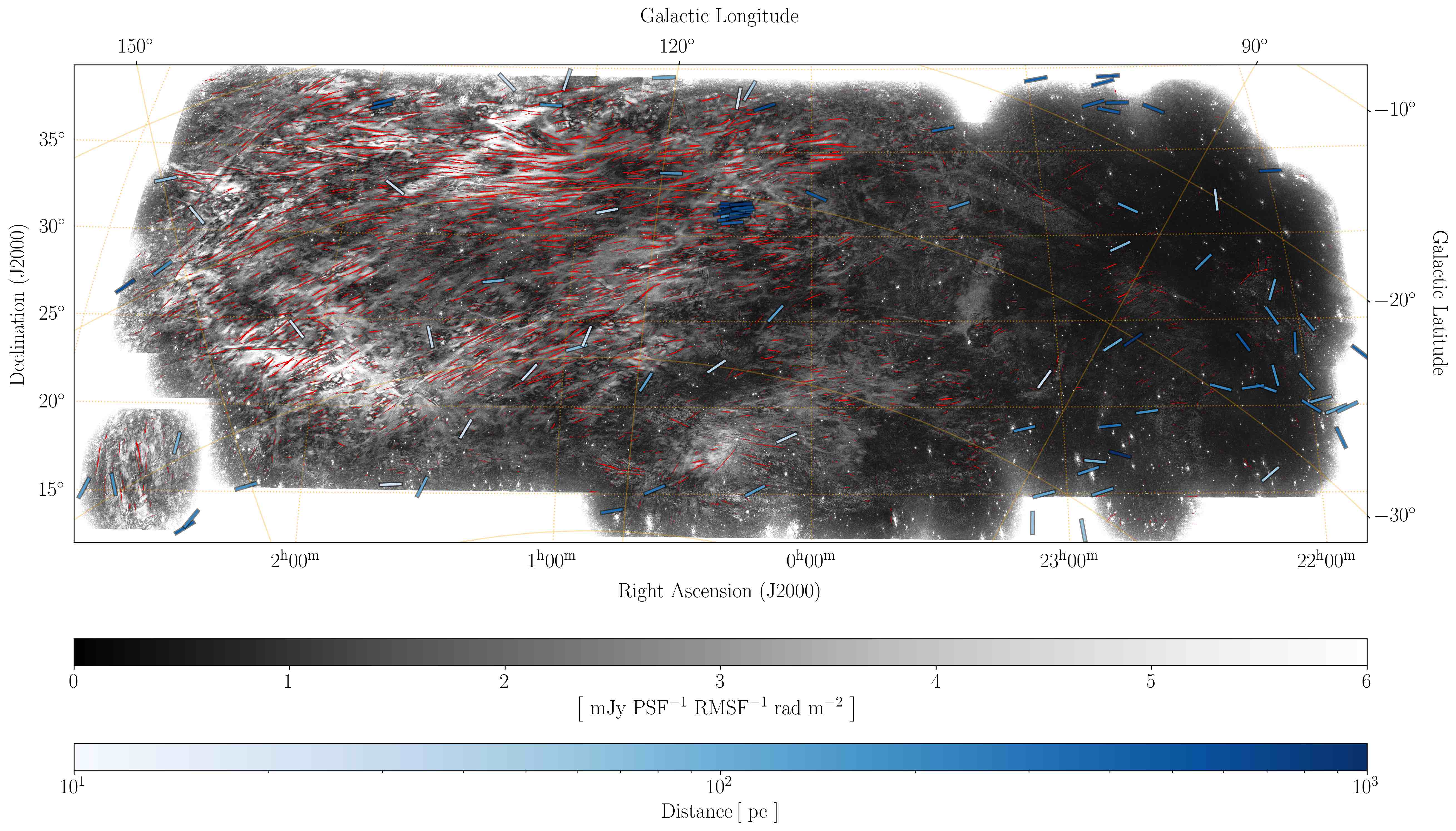}
      \caption{Maximum polarised intensity image overlaid with starlight polarisation measurements in blue and depolarisation canals detected by RHT in red.}
         \label{RHT+stars}
\end{figure*}

\begin{figure*}
  \centering
    \includegraphics[width=0.9\textwidth]{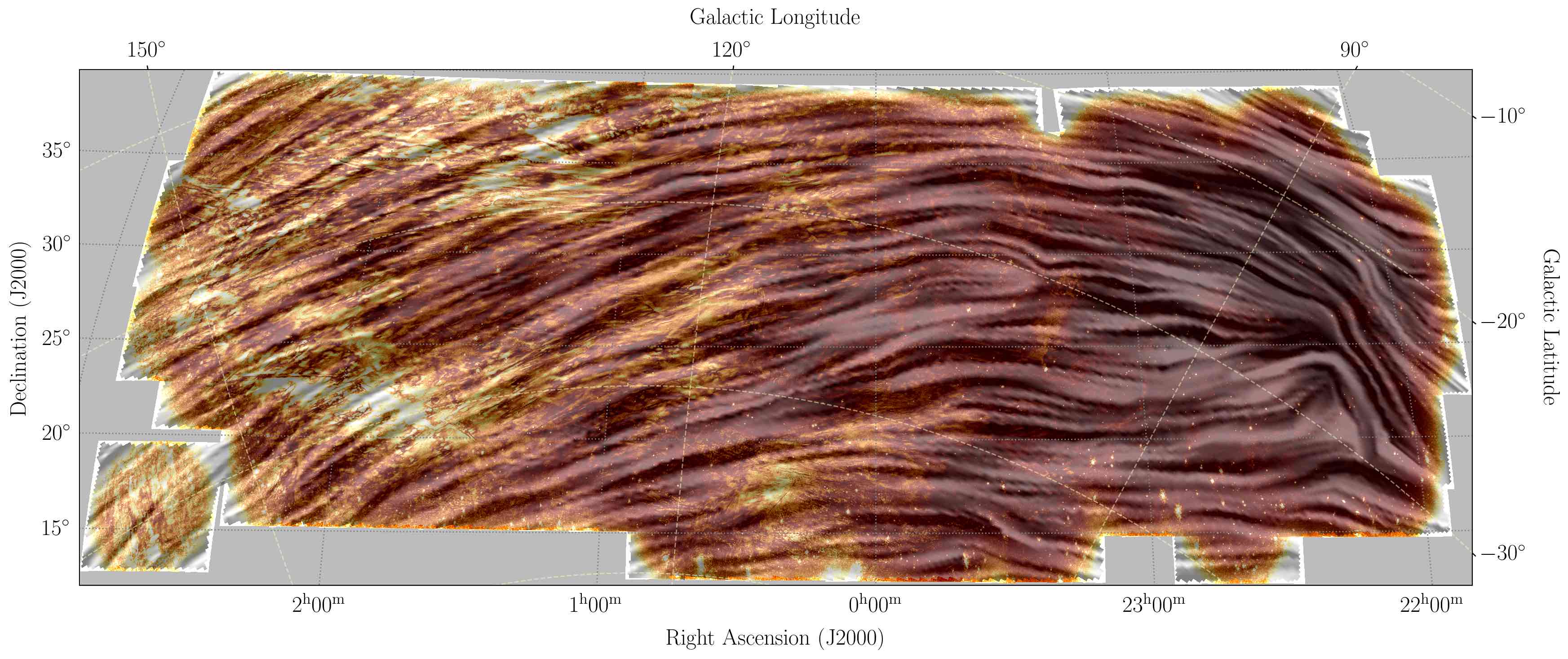}
  \caption{Orientation of the \textit{Planck} 353 GHz plane-of-sky magnetic field, represented by a drapery pattern overlaid on the image of maximum polarised intensity (orange). }
  \label{LIC}
\end{figure*}

In this section, we quantify the alignment between depolarisation canals and the Galactic magnetic field, starlight polarisation and HI filaments. We use the alignment between canals and starlight polarisation to estimate the distance to Faraday structures.

\subsection{Detecting depolarisation canals}
We perform RHT on the inverted maximum polarised intensity image to identify linear depolarisation canals. We used the reciprocal inversion, as it highlights canals while suppressing the surrounding emission. The choice of RHT parameters is the same as in \citet{jelic18}, \citet{turic21} and \citet{erceg24}, ($D_K$, $D_W$, $Z$) = (10, 61, 0.8), which at LoTSS vlow resolution corresponds to ($D_K$, $D_W$, $Z$) = (8 arcmin, 50 arcmin, 0.8). The backprojection of RHT is drawn with red lines in Fig. \ref{RHT+stars}. Identified canals have a typical width of the PSF, length of several degrees, and follow the morphology of polarised intensity structures in the area east from $\rm{23^h15^m}$. On the contrary, the emission west from $\rm{23^h15^m}$ is depolarised and the canals are sparse.

\subsection{Magnetic field}\label{subsec:magfield}
We reconstructed the plane-of-sky magnetic field lines from the Planck 353 GHz observations (Fig. \ref{LIC}). The magnetic field is ordered across a large part of the mosaic, with slight deviations in the westernmost area. 

Using the same approach as in \citet{erceg24} we divided the area into small rectangular patches with sides measuring 1.5 degrees. Within each patch, we computed the circular mean of both the magnetic field and depolarisation canal orientation. The difference between the two is the $\Delta \psi$ in the calculation of PRS. The obtained PRS is $7.4~\pm~0.5$, which surpasses the significance threshold of 2.7. The depolarisation canals align well with the direction of the plane-of-sky magnetic field and we assume that Faraday structures associated with them occupy the same volume of space along the LOS. In the next subsection, we utilise starlight polarisation to estimate the distance to these structures.

\subsection{Starlight polarisation}

\begin{figure}
   \centering
   \includegraphics[width=\hsize]{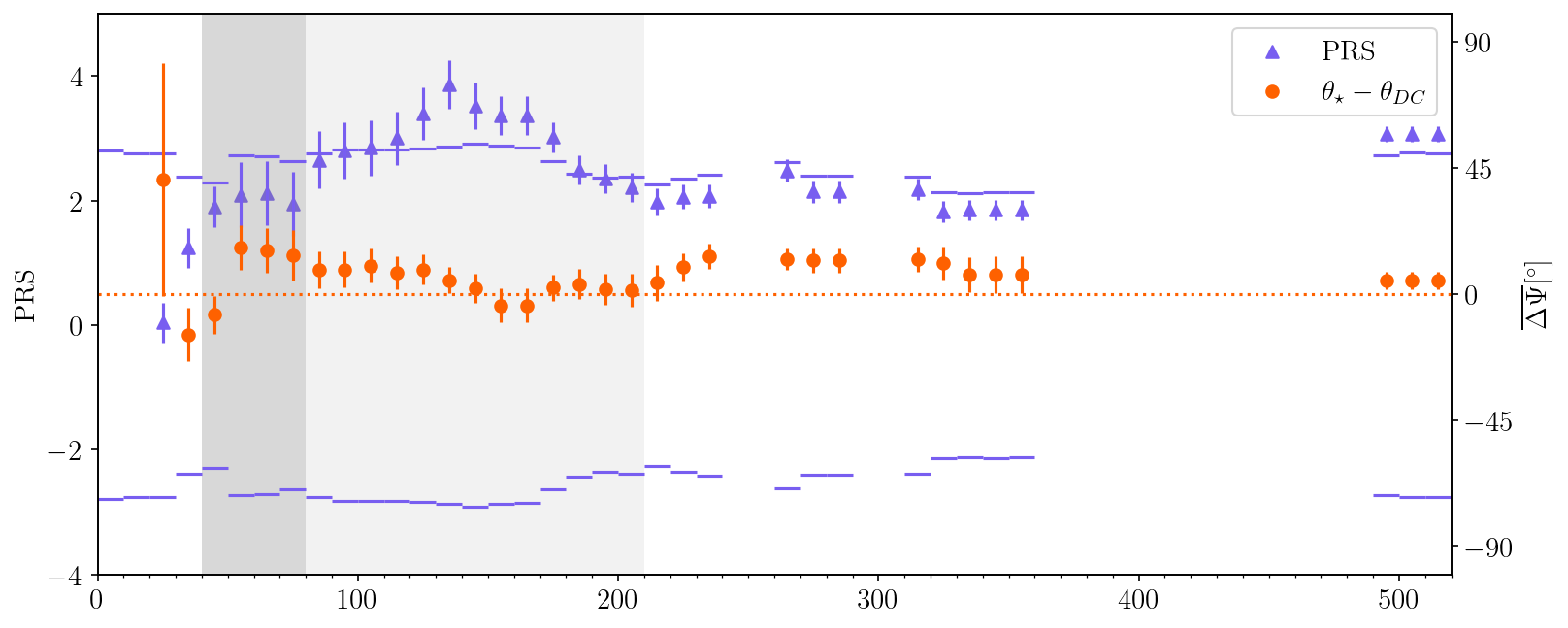}
      \caption{PRS values (purple triangles), their errors and the
significance threshold (horizontal bars) computed between starlight polarisation and depolarisation canals in the region east from $RA~\mathrm{23^h15^m}$. Orange marks represent $\Delta \psi$ and the associated error. The result for each bin is represented by a point at its lower border. The lighter shaded area represents all bins in which we find alignment, while the darker shaded area marks the estimated range for the minimum distance to the structures.}
    \label{PRS_L}
\end{figure}

 In the \citet{panopoulou23} catalogue we found 117 stars in our region. Around half of those stars are located in the western part, where polarised emission is low and there are no well-defined depolarisation canals (Fig. \ref{RHT+stars}). For this reason, we focused on the area east from $\rm{23^h15^m}$, where we found 64 stars. We performed stellar tomography by separating the stars in distance bins 100 pc wide, each separated by 10 pc, so that the first few bins are $0-100~\mathrm{pc}$, $10-110~\mathrm{pc}$, etc. With this binning, the average sample size is 9 stars per bin. 
 
 We search for alignment between depolarisation canals and the starlight polarisation. We compute the PRS, its error and the significance threshold and present the results in Fig. \ref{PRS_L}, where the PRS result for each distance bin is plotted at the lower border of the bin. The PRS analysis reveals alignment for stars beyond 80 pc. 
 
 To determine the lower and the upper bound of the minimum distance we used the criteria outlined in \citet{panopoulou21}. Each bin for which the PRS value is above the significance threshold and $\Delta \psi \leq 20^\circ$ qualifies as an upper bound. If both criteria are met consistently across multiple bins, then the bin at the largest distance is chosen as the upper bound. Lower bound is defined as a bin where $\Delta \psi$ abruptly drops to $\leq 20^\circ$. We established the lower and the upper bounds to the minimum distance as 40 and 80 pc, respectively (Fig. \ref{PRS_L}). The bin size defines an error margin of 100 pc for these estimates. It should be noted that the spatial distribution of available starlight polarisation measurements may have impacted the result. Most of the stars closer than 100 pc are found in the eastern area, while most of the stars in the central area are much farther and concentrated around $RA$, $Dec$ = ($\mathrm{0^h20^m}$, $31\degree$), as can be seen in Fig. \ref{RHT+stars}.

\subsection{HI filaments}

\begin{figure*}
   \centering
   \includegraphics[width=0.97\textwidth]{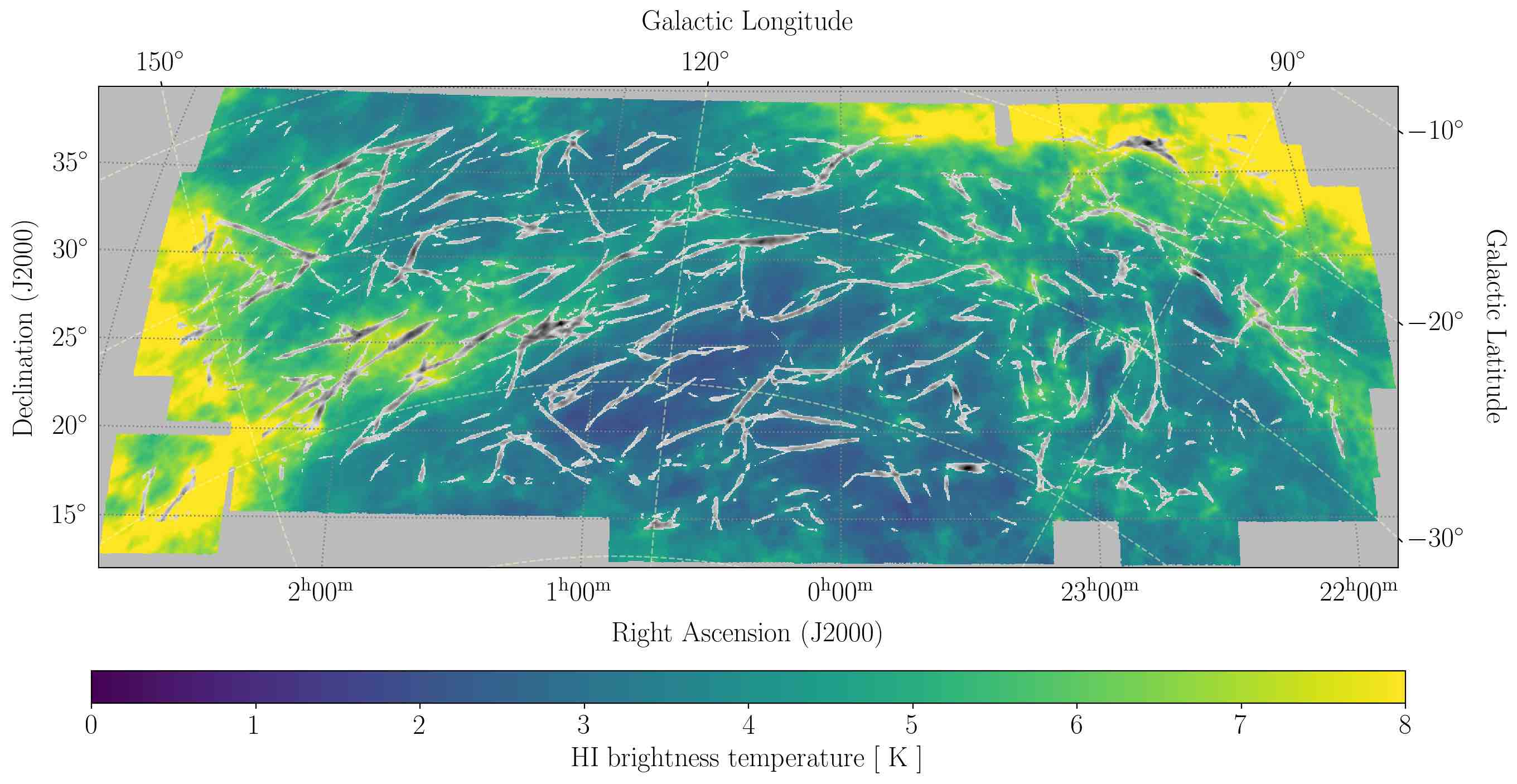}
      \caption{Integrated HI intensity in Kelvins, overlaid with RHT backprojection (grey lines). }
         \label{hi}
\end{figure*}
We search for correlation with HI filaments found in the image of HI temperature brightness integrated over $v_\mathrm{LSR} \leq 30~\mathrm{km~s^{-1}}$ (Fig. \ref{hi}). We quantified the orientation of filaments by performing an RHT analysis with parameters $D_K$, $D_W$, $Z$ = (55, 10, 0.7). As in Sect. \ref{subsec:magfield}, we separated the data set in square patches with sides of length 1.5 degrees and for each calculated the mean circular orientation of filaments and the depolarisation canals. The two sets of angles are used to compute $\Delta \psi$. The two tracers show a strong correlation with $\mathrm{PRS} =14.1\pm0.4$ and the significance threshold of 3.1. This alignment is found over a large velocity range indicating an ordered magnetic field along the LOS \citep{clark18, clark19}.

\section{Connection to the Local Bubble}
\label{sec:discussion}
\begin{figure}
   \centering
   \includegraphics[width=\hsize]{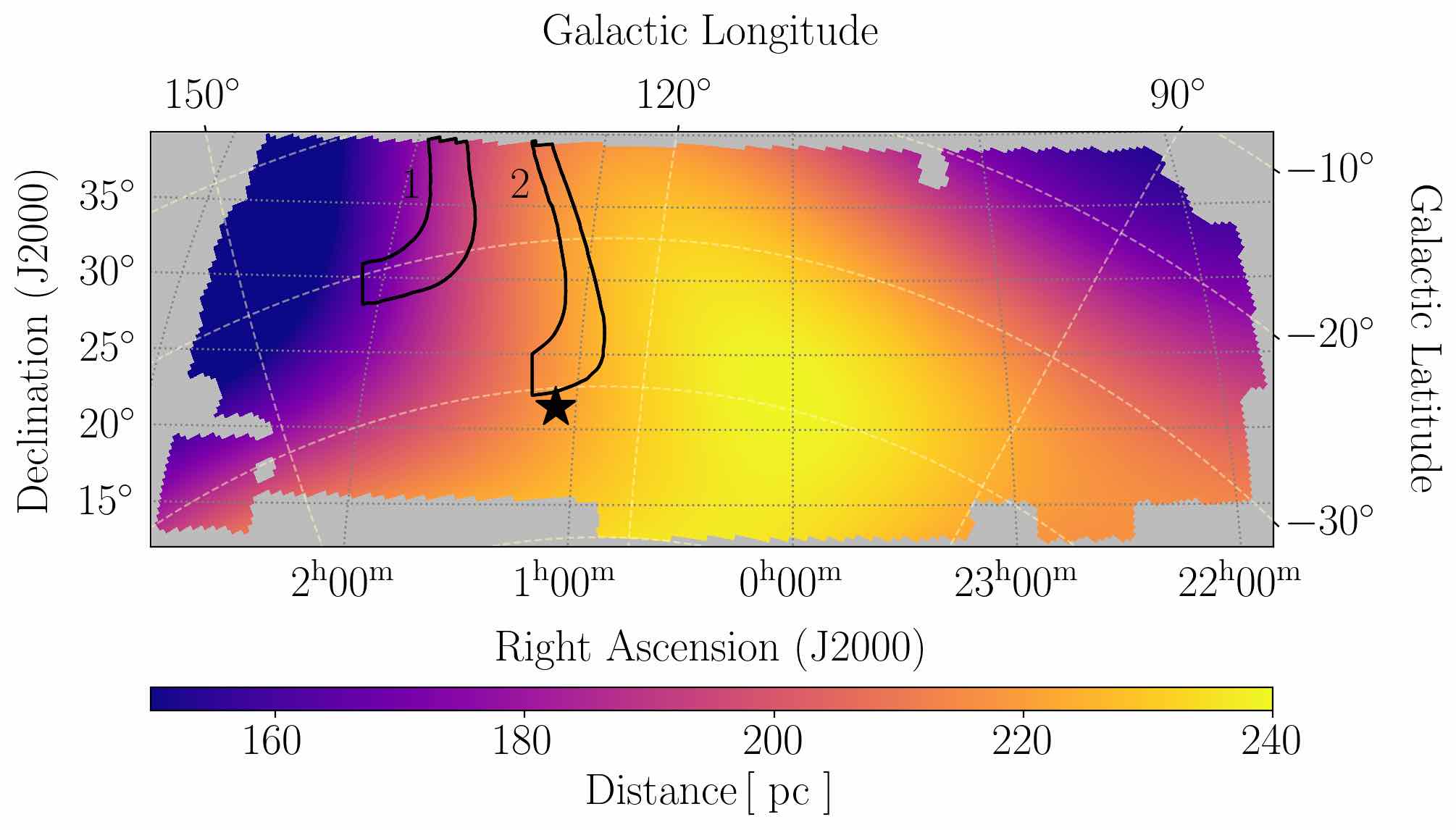}
      \caption{Distance to the Local Bubble from \citet{pelgrims20}. Black contours mark two regions that we compare in Sect. \ref{subsect:LBwall}. The black star represents the location of binary star 74 Psc A and B.}
         \label{LB_dist}
\end{figure}

\begin{figure}
   \centering
   \includegraphics[width=\hsize]{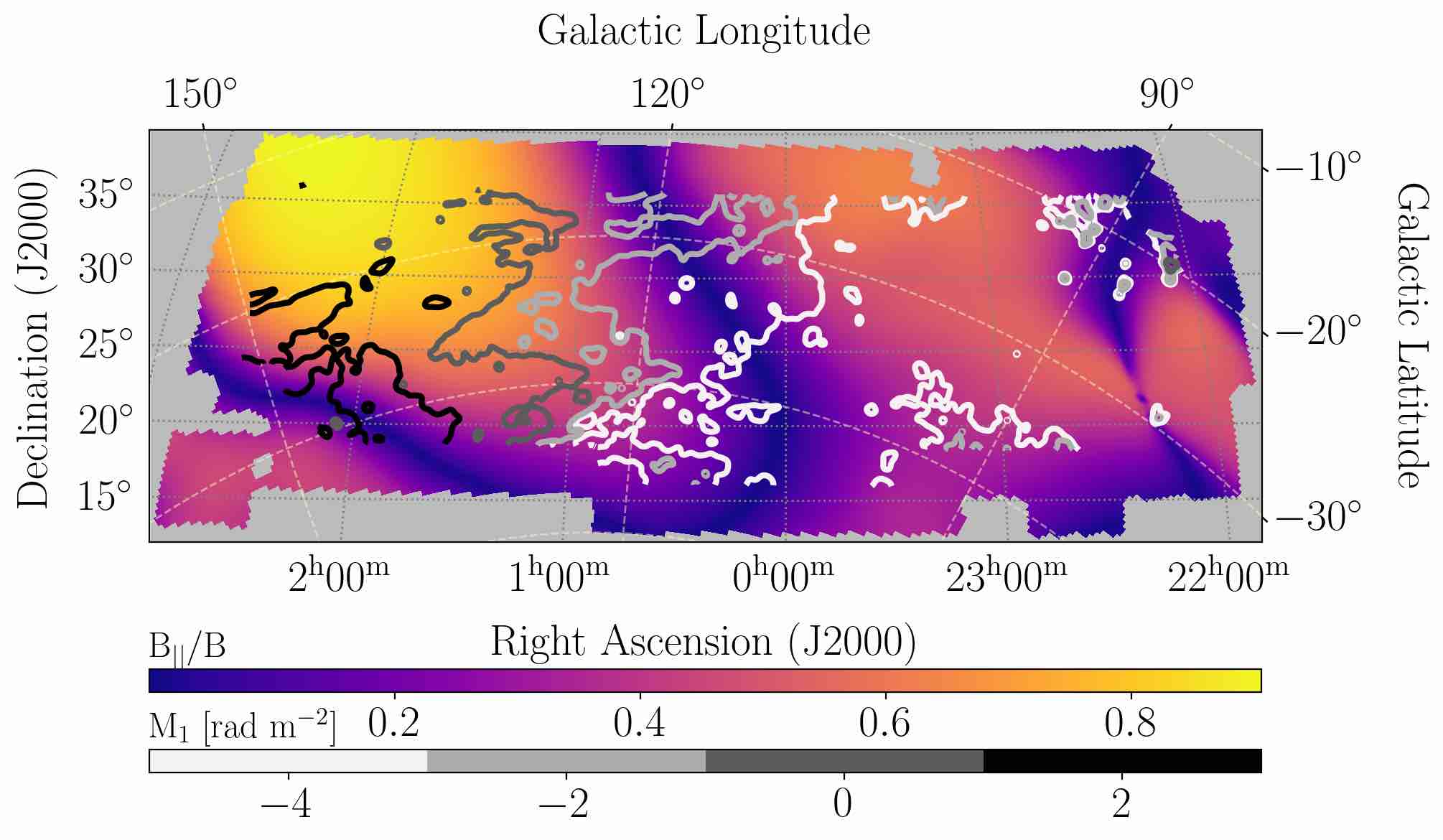}
      \caption{LOS magnetic field component in the wall of the Local Bubble, represented through fraction $B_\mathrm{||}/B$, extracted from \citet{pelgrims20}. The contours trace different levels from the M1 map.}
         \label{B_LOS}
\end{figure}

\begin{figure}
   \centering
   \includegraphics[width=\hsize]{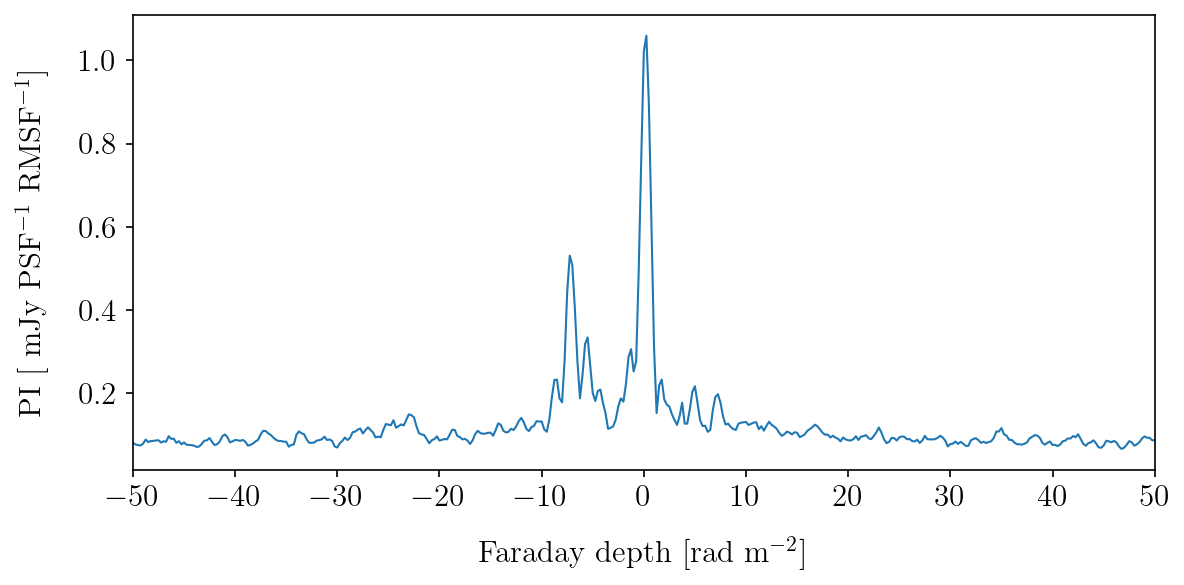}
      \caption{Faraday spectrum at the location of stars Psc A and B.}
         \label{PscAB_spec}
\end{figure}

\begin{figure}
   \centering
   \includegraphics[width=\hsize]{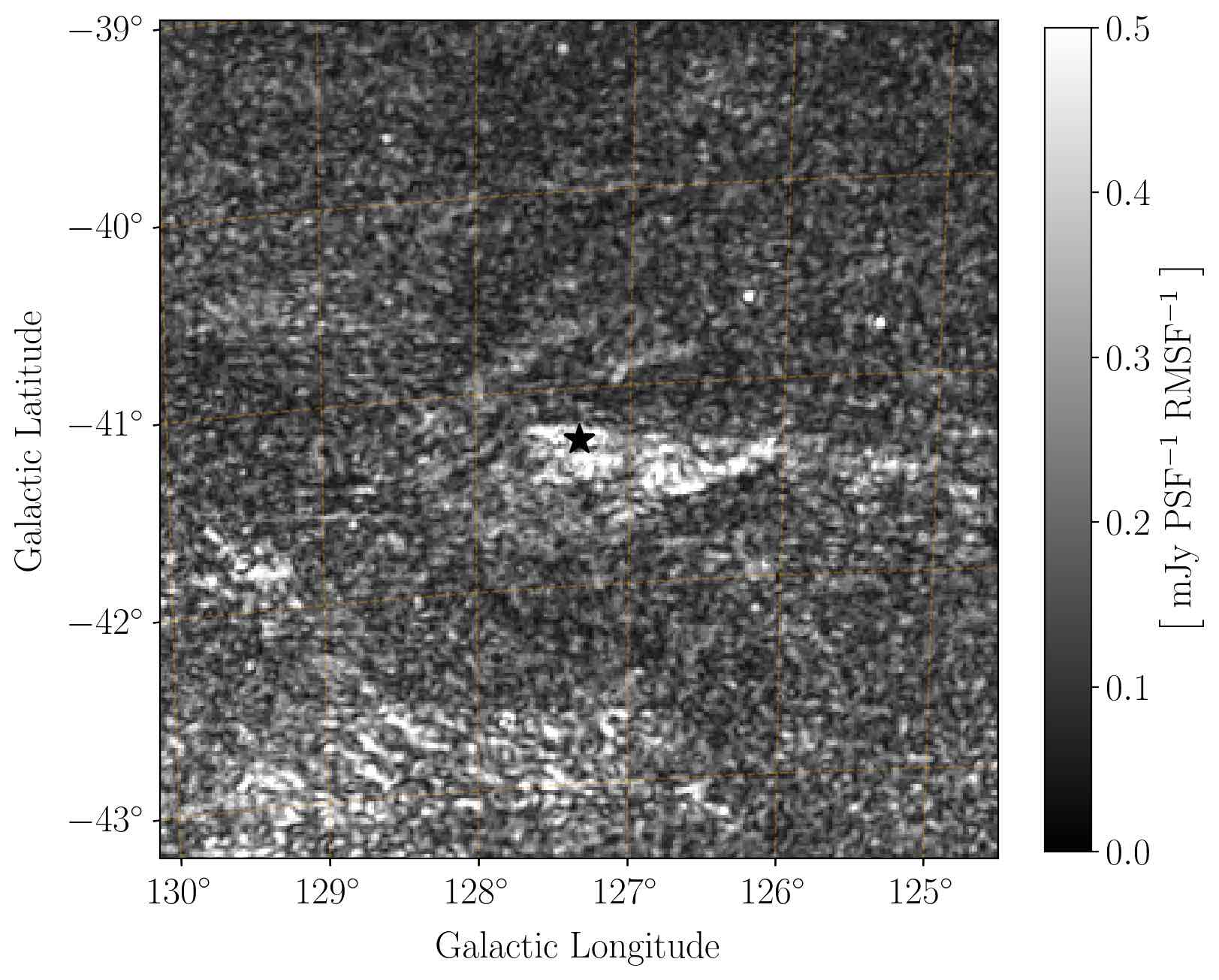}
      \caption{ Close up view on the area around the binary star Psc A and B (black star). The intensity is averaged over a range from -8 to $-6~\mathrm{rad~m^{-2}}$. }
         \label{zoom_PscAB}
\end{figure}

\begin{figure}
   \centering
   \includegraphics[width=\hsize]{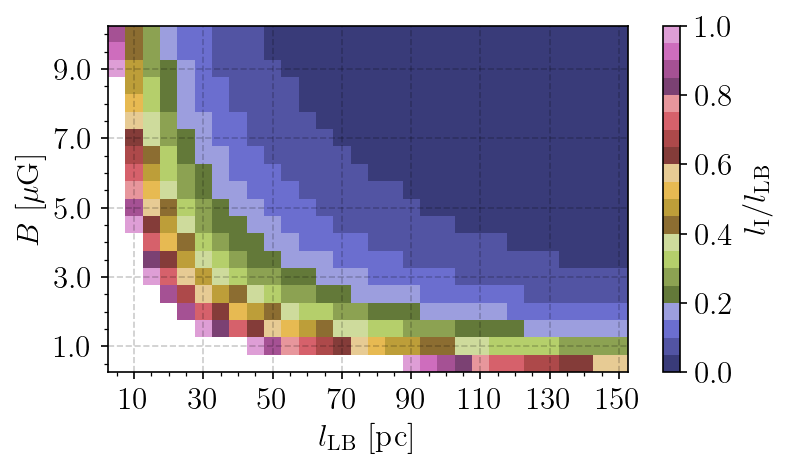}
      \caption{Ionised to total Local Bubble wall thickness as a function of the magnetic field strength and the total Local Bubble wall thickness. Values higher than 1 are masked, as they are not physical in this toy model.}
         \label{frac_li}
\end{figure}

\begin{figure}
   \centering
   \includegraphics[width=\hsize]{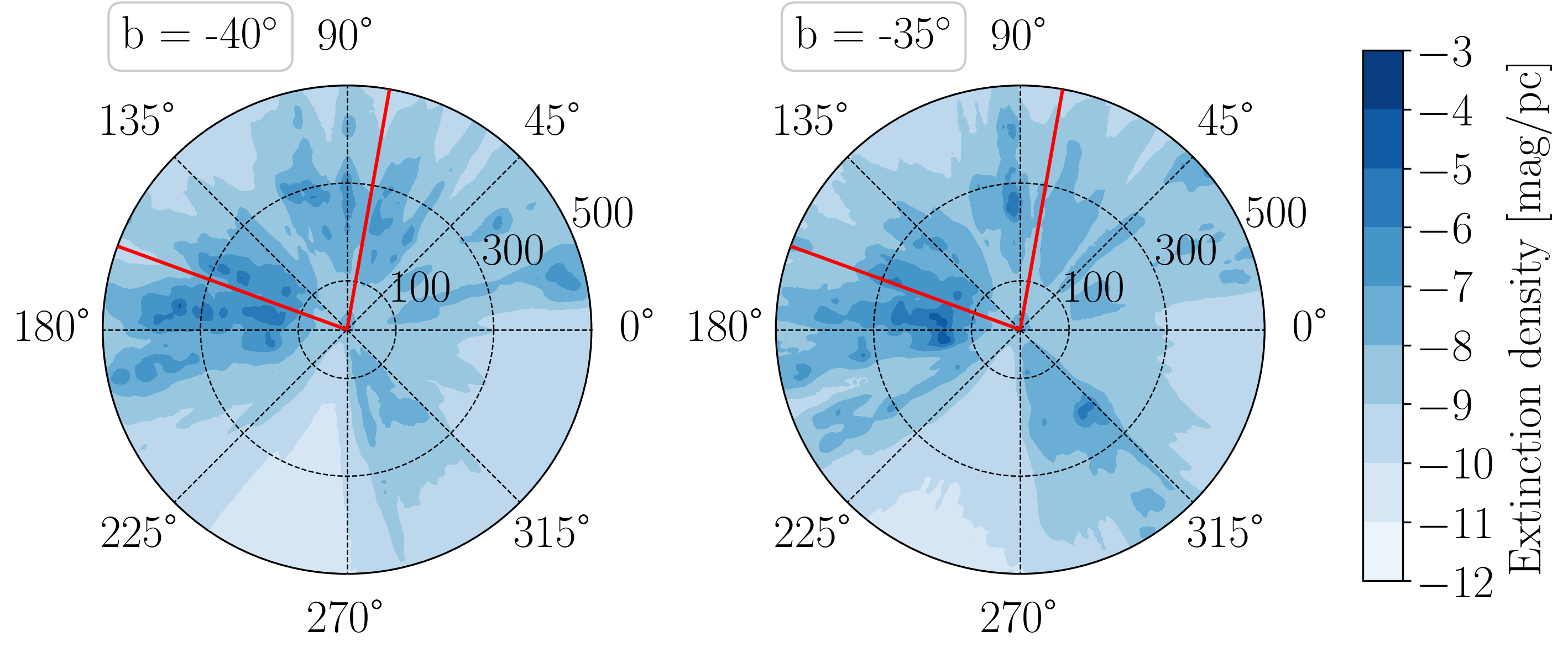}
      \caption{Dust extinction at different Galactic latitudes, made with data of \citet{vergely22}. In each panel, the Sun is in the centre and the circular maps represent a 2D projection of a conical cross-section along a constant Galactic latitude (corresponding to geometry of a martini glass). Galactic longitudes are denoted on the angular coordinate and the distance to the Sun is the radial coordinate. The two red lines bound the area covered by the mosaic. }
         \label{vergely}
\end{figure}

We examined the \citet{pelgrims20} map of distances to the inner Local Bubble wall in the whole area of the mosaic, produced by expanding the wall in spherical harmonics with $l_\mathrm{max}=10$ (Fig. \ref{LB_dist}). From this map, we computed the mean distance to the Local Bubble wall to be 205 pc, beyond our estimate for the minimum distance to the depolarisation structures.

The map of distances in Fig. \ref{LB_dist} reveals the concave shape of the Local Bubble wall - the wall is farthest near the centre of the mosaic and it is closer at the sides. Assuming a constant magnetic field that is tangential to the wall \citep[which may not hold everywhere on the Local Bubble, see][]{halal24}, we employed the \citet{pelgrims20} map of the Local Bubble wall inclination to deduce the fraction of the magnetic field along the LOS (Fig. \ref{B_LOS}). Following the concavity of the Local Bubble wall, the magnetic field is almost completely in the plane-of-sky in the middle of the image and then gradually transitions towards LOS on the sides. The gradient in Faraday depths observed in the LoTSS data follows a similar pattern, as it goes from positive to negative Faraday depths when observed left to right and crosses $0~\mathrm{rad~m^{-2}}$ in the central area of the mosaic. We conclude that the gradient is tracing the concavity of the Local Bubble wall.

\subsection{Implications on the ISM in the Local Bubble cavity}
\label{subsect:Disc1}
If we assume that all of the polarised emission observed in LoTSS comes from the Local Bubble wall, then the location of emission at $0~\mathrm{rad~m^{-2}}$ should coincide with the area where the wall lies entirely in the plane-of-sky, i.e. $B_{||}/B \sim 0$. Contours of emission at different Faraday depths (Fig. \ref{B_LOS}) reveal a significant difference between the two. On smaller scales, this discrepancy is caused by the difference between the resolution of LoTSS data and the model. However, there is also a negative global offset of several $\mathrm{rad~m^{-2}}$. In this subsection, we explore the possibility that this excess Faraday rotation could have been caused by the medium inside of the Local Bubble cavity. To do this, we use a simplification of Eq. \ref{eq:FRangle}, where we assume that none of the parameters change with d$l$.

The average parallel component of the magnetic field, $\langle B_{||}\rangle$, can be estimated from measurements of pulsar RM and dispersion measure (DM) as $1.232\cdot RM/DM$ \citep[e.g.][]{xuhan19}. Since there are no observed pulsars in our field, we focused on all the pulsars in the first 200 pc with RM measurements. In the \citet{manchester05}\footnote{ATNF Pulsar Catalogue, \url{https://www.atnf.csiro.au/research/pulsar/psrcat/}} catalogue we found 16 such pulsars and calculate that, on average, $\langle B_{||}\rangle=1~\mathrm{\mu G}$. Although some of these pulsars are within the Local Bubble cavity and some may be beyond the wall, the range of values for $\langle B_{||}\rangle$ is not large (most of them fall within 0.5 to $2~\mathrm{\mu G}$). Thus, throughout the discussion, we adopt this roughly estimated value as the average for both the cavity and the wall.

For an estimate of electron density in the Local Bubble cavity we turn to 
\citet{redfield08}, who use spectroscopic observations of stars closer than 100 pc to study the CLIC. Two of the stars they consider (74 Psc A and B) are found within our field of view, at galactic coordinates ($l$,$b$) = ($127.34\degree$, $-41.28\degree$), with a mean distance of 70.7 pc\footnote{The two stars form a binary, so we take the average of all measurements.} (well within the Local Bubble cavity, Fig. \ref{LB_dist}). By analysing the $\mathrm{C_{II}}$ fine-structure excitation the authors found two different clouds along this LOS, with average free electron densities of $n_{e,1} = 0.04~\mathrm{cm^{-3}}$ and $n_{e,2,\mathrm{LIC}} = 0.6~\mathrm{cm^{-3}}$. The authors noted that the denser component is most likely associated with the LIC. We focus the discussion in this subsection on the LOS towards Psc A and B.

The Faraday spectrum in this direction is two-peaked, with peaks around -7 and $0~\mathrm{rad~m^{-2}}$ (Fig. \ref{PscAB_spec}). While both peaks are a part of diffuse emission, the one around zero agrees well with the observed gradient emission and seems to be associated with the curved Local Bubble wall. However, the negative peak stands out from the gradient. At $-7~\mathrm{rad~m^{-2}}$ the bulk of the gradient emission is found around $l=100\degree$, around $30\degree$ west from the location of Psc A and B. The emission towards Psc A and B is part of a smaller, patchy structure, which is not connected to the gradient and can be seen in Fig. \ref{zoom_PscAB}. We averaged the emission shown in Fig. \ref{zoom_PscAB} over a narrow Faraday depth range, spanning from $-8~\mathrm{rad~m^{-2}}$ to $-6~\mathrm{rad~m^{-2}}$. This allowed us to capture Faraday structures with similar properties, given the $1~\mathrm{rad~m^{-2}}$ resolution in Faraday space. If we assume that the magnetic field in the cavity is uniform and oriented towards us (as is indicated by the global negative offset), we may associate the negative peak in the Faraday spectrum towards Psc A and B with the contribution of the medium inside the cavity. 

If we assume that the two clouds found by \citet{redfield08} fill out the full LOS towards Psc A and B and that their combined effect on the synchrotron emission is the Faraday peak at $-7~\mathrm{rad~m^{-2}}$, we can calculate that their expected dimensions along the LOS should be $l_1 = 59.7~\mathrm{pc}$ and $l_{2, \mathrm{LIC}} = 10~\mathrm{pc}$, respectively. With these assumptions, the warm partially ionised LIC would fill up 15\% of the volume towards the Local Bubble along this particular LOS.

Other combinations of values become possible if we relax the assumption of the uniform magnetic field inside the cavity. Nevertheless, this simple calculation suggests that LoTSS may not be observing just the Local Bubble wall but also the signature of the warm ionised clouds within the Local Bubble cavity.

\subsection{Implications on the structure of the Local Bubble wall}
\label{subsect:LBwall}
Given the shared geometry and comparable distance of structures associated with the depolarisation canals and the Local Bubble wall, we now include the wall into consideration. We select two areas where the gradient in Faraday depths follows the gradient of $B_\mathrm{||}$ created by the curvature of the wall (marked in Fig. \ref{LB_dist}). By confining our analysis to these two areas, we aim to minimise the interference from foreground and/or background Faraday structures that could be affecting the result. We assume that the morphological similarities between the two gradients indicate comparable conditions in the foreground/background ISM, whose effect cancels out in an analysis which considers only the relative change between the two areas. Once more, we use Eq. \ref{eq:FRangle} and investigate which conditions in the Local Bubble wall can produce the observed difference in mean Faraday depth, represented by M1.

\citet{pelgrims20} estimate the wall's thickness to be between 50 and 150 pc; however, this should be taken with care as the authors note that the exact outer limit of the wall is difficult to determine. As heightened dust extinction, which traces the Local Bubble wall, is connected with the neutral medium, we consider a combination of cold and warm neutral medium and for the density of free electrons take $n_{e,\mathrm{WNM+CNM}}=0.008~\mathrm{cm^{-3}}$ \citep{heileshaverkorn12}. From the M1 map, we calculate $\Delta M1 = 1.9~\mathrm{rad~m^{-2}}$ as the difference in M1 average in the two chosen areas, and from the map of $B_\mathrm{||}$ we calculate the mean $\Delta (B_\mathrm{||}/B) = 0.3$. To recover the observed $\Delta M1$ with conditions listed above, we would require the total magnetic field strength of 20 to $7~\mathrm{\mu G}$, for the wall thickness of 50 to 150 pc respectively. In the context of total magnetic field estimates in the warm clouds found in the cavity of the Local Bubble of 3 to $5~\mathrm{\mu G}$ \citep{schwadron19, snowden14} and our estimate of the average parallel component of the magnetic field along the LOS toward nearby pulsars of $1~\mathrm{\mu G}$, these values appear to be too high. However, it is possible that the magnetic field in the wall was enhanced due to compression.

We consider a different approach to the problem. If we set $B_\mathrm{||} = 1~\mathrm{\mu G}$, with other parameters remaining as in the text above, we obtain Faraday depth of 0.3 to $1~\mathrm{rad~m^{-2}}$, for wall thickness of 50 to 150 pc, respectively. The missing Faraday rotation may be produced by an ionised front created through the interaction of the neutral wall with the ionised medium on the near or far end of the Local Bubble wall. We took the mean of free electron density in a combination of the warm ionised medium and the warm partially ionised medium to be $n_{e,\mathrm{WIM+WPIM}}=0.18~\mathrm{cm^{-3}}$ \citep{heileshaverkorn12}. In this toy model, we assumed that the Local Bubble wall consists of two components, a warm ionised front of thickness $l_\mathrm{I}$ and a neutral wall of thickness $l_\mathrm{N}$ and that the Faraday rotation of the two components is additive. With the wall thickness of $l_\mathrm{LB} = l_\mathrm{I}+l_\mathrm{N}=50 - 150~\mathrm{pc}$ we need $l_\mathrm{I} = 43 - 39~\mathrm{pc}$ to account for the observed relative difference in the Faraday depth. This would imply that the fraction of ionised to total path length in the Local Bubble wall is between 86\% and 26\%. High values of this fraction point to a thin neutral wall border, while values higher than one would imply the presence of a tunnel or a chimney in the Local Bubble wall. 

Combining the two ideas - allowing for the total magnetic field strength of up to $10~\mathrm{\mu G}$  and for the existence of an ionised front of finite thickness, we produce a range of values for the fraction of ionised to total path length (Fig. \ref{frac_li}). Weaker magnetic field and a thinner wall, which are closer to what is expected from the observations, seem to produce high values of the fraction, implying a very thin component of the cold neutral medium accompanied by a thick ionised front in the Local Bubble wall. 

We examine this result further by observing dust extinction maps of \citet{vergely22} in Fig. \ref{vergely}. The maps show a presence of a lower extinction density tunnel-like feature at latitudes -35 and -40 and longitudes between 110 to 130 degrees, which could be a cause of a bias toward higher fractions in plot in Fig. \ref{frac_li}.

\section{Limitations of LoTSS data}
\label{sec:limitations}
The total rotation measure produced by our Galaxy may be obtained from the Faraday Sky map \citep{hutschenreuter22}, which was constructed using RM of extragalactic radio sources, including the ones from the LoTSS RM grid \citep{osullivan23}. In the area corresponding to the LoTSS mosaic, the values in the Faraday Sky map are predominantly negative, with an average around $-67~\mathrm{rad~m^{-2}}$. These values are much higher than the values in the M1 map. Additionally, the morphology in Faraday Sky does not in any way correlate to the morphology of the LoTSS data, indicating that there is a large portion of volume that is not observed with LoTSS. If we limit the analysis to a more local ISM by examining the RM of a sample of pulsars from the ATNF catalogue discussed in Sect. \ref{subsect:Disc1}, we still find discrepancies. The absolute RM of all pulsars within 200 pc is on average $26.3~\mathrm{rad~m^{-2}}$, which again greatly exceeds what is observed by LoTSS. The missing rotation measure may in both cases be attributed to different limits of the low radio frequency interferometric observations. 

Firstly, due to the high sensitivity of low-frequency synchrotron emission to Faraday rotation, the polarised emission depolarises along relatively short LOS through the effect of depth depolarisation. The remaining polarisation fraction of synchrotron emission in LOFAR data was estimated to be only about $1-4~\%$ \citep{iacobelli13, jelic14, jelic15, turic21}.

Another cause may be the lack of sensitivity to Faraday thick structures, which contribute to the total RM but cannot be reconstructed using RM-synthesis at these frequencies. Modified Fourier transform used in the RM-synthesis may reconstruct only the edges of Faraday thick structures and only at a fraction of their true intensity \citep[around 12\%,][]{vaneck17}. The amount of emission we're missing due to Faraday thickness could be estimated by analysing Faraday data at higher frequencies, such as radio polarimetric data at 1.4 GHz observed by the DRAO Global Magneto-Ionic Medium Survey \citep[GMIMS,][]{wolleben21}. This data has an angular resolution of 40 arcmin and a Faraday depth resolution of $140~\mathrm{rad~m^{-2}}$. This means that DRAO GMIMS can observe structures that are Faraday thick for LoTSS but, due to beam depolarisation, has lower sensitivity and may be missing some of the faint nearby structures.
Additionally, Faraday rotation at 1.4 GHz is much less pronounced, which diminishes the effect of depth depolarisation, making the observed LOS longer and foreground structures harder to isolate. Considering this, we compared the LoTSS M1 map with the DRAO GMIMS M1 map \citep{dickey19}. The two maps showed no clear morphological correlation. The mean $M1_\mathrm{DRAO-GMIMS}$ is around 4 times larger than the $M1_\mathrm{LoTSS}$, making it difficult to estimate the shared observed volume between the two datasets and therefore also the emission that LoTSS might be missing due to Faraday thickness. However, comparing the LoTSS data with data at intermediate frequencies, such as the upcoming Canadian Hydrogen Intensity Mapping Experiment \citep[CHIME,][]{chime22}\footnote{\url{https://chime-experiment.ca/en}} data may help bridge this gap in the near future.

Finally, LoTSS observations are missing short baselines, which makes them sensitive only to structures on scales smaller than around 1 degree. A structure such as the Local Bubble, may have a constant, large-scale contribution to the RM that is filtered out. Mapping the missing large-scale Galactic emission will be possible with future LOFAR observations, such as the AARTFAAC \footnote{\url{http://aartfaac.org/index.html}} project (Amsterdam - ASTRON Radio Transients Facility and Analysis Centre), which will provide almost full $uv$ coverage at short-baselines.

\section{Summary and conclusion}\label{sec:summary&conclusion}

With this paper, we presented the Faraday cube of the LoTSS observations in the high-latitude inner Galaxy. The data revealed organised morphology of polarised emission accompanied by straight and narrow depolarisation canals at $RA$ east from $\mathrm{23^h15^m}$ and depolarised, patchy emission in the western area. The emission in the east is characterised by a large gradient, roughly perpendicular to the orientation of depolarisation canals, spanning Faraday depth range from -15 in the central part of the mosaic to +8 $\mathrm{rad~m^{-2}}$ in the eastern part. By inspecting the Faraday cube and calculating Faraday moments, we concluded that most of the polarised emission has a single peaked Faraday spectrum, i.e. comes from a single dominating structure. 

We found alignment between depolarisation canals and the \textit{Planck} plane-of-sky magnetic field, indicating that the traced structures are mutually connected and located at a similar distance from the Sun. Significant alignment is also noted between depolarisation canals and filaments present in the integrated HI image. Utilising starlight polarisation measurements, we estimated the distance to the structures associated with depolarisation canals, determining a minimum distance range of 40 to 80 pc with an uncertainty of 100 pc. This estimate puts the origin of the Faraday structures in the proximity of the Local Bubble wall.

We compared our data with the \citet{pelgrims20} model of the Local Bubble. The gradient we observed matches the expected curvature of the Local Bubble in the studied area within several $\mathrm{rad~m^{-2}}$. We show that this difference may be produced by Faraday rotation caused by the warm clouds within the Local Bubble's cavity. We extended our analysis to the wall of the Local Bubble and showed that the curvature expected from the \citet{pelgrims20} model combined with expected values of magnetic field and free electron density in the neutral medium is not enough to reproduce the gradient in Faraday depths we observe. We suggested a toy model of the Local Bubble wall that includes two types of medium, an ionised front and a cold shell. By varying the total magnetic field from 1 to $10~\mathrm{\mu G}$ and the total wall thickness from 5 to 150 pc, we obtained different values for the fraction of ionised front thickness in the total wall thickness. 

Finally, we discussed the limitations of LoTSS data in terms of strong depolarisation at low radio frequencies, lack of sensitivity to Faraday thick structures and missing large scales. We compared our data to the DRAO GMIMS data set but did not find a clear correlation that could help us estimate the amount of missing emission. Observations from upcoming surveys may help us better understand the limitations of Faraday tomography in different frequency regimes and shed light on the complex nature of ISM.

\begin{acknowledgements}
We are thankful to A. Bracco and R. Benjamin for useful discussions related to this work. A.E. and V.J. acknowledge support from the Croatian Science Foundation for project IP-2018-01-2889 (LowFreqCRO). M.H. acknowledges funding from the European Research Council (ERC) under the European Union's Horizon 2020 research and innovation programme (grant agreement No 772663). M.J.H. acknowledges support from the UK STFC [ST/V000624/1]. L.G. acknowledges funding by the Deutsche Forschungsgemeinschaft (DFG, German Research Foundation) under Germany's Excellence Strategy -- EXC 2121 ``Quantum Universe'' -- 390833306. 
The authors acknowledge Interstellar Institute's program "II6" and the Paris-Saclay University's Institut Pascal for hosting discussions that nourished the development of the ideas behind this work. This paper is based on data obtained with the International LOFAR Telescope (ILT) under project codes $\rm{LC2\_038}$ and $\rm{LC3\_008}$. LOFAR (van Haarlem et al. 2013) is the Low-Frequency Array designed and constructed by ASTRON. It has observing, data processing, and data storage facilities in several countries, that are owned by various parties (each with their funding sources), and that are collectively operated by the ILT foundation under a joint scientific policy. The ILT resources have benefitted from the following recent major funding sources: CNRS-INSU, Observatoire de Paris and Université d'Orléans, France; BMBF, MIWF-NRW, MPG, Germany; Science Foundation Ireland (SFI), Department of Business, Enterprise and Innovation (DBEI), Ireland; NWO, The Netherlands; The Science and Technology Facilities Council, UK; Ministry of Science and Higher Education, Poland. This research made use of Montage. It is funded by the National Science Foundation under Grant Number ACI-1440620 and was previously funded by the National Aeronautics and Space Administration's Earth Science Technology Office, Computation Technologies Project, under Cooperative Agreement Number NCC5-626 between NASA and the California Institute of Technology.
\end{acknowledgements}

\bibliographystyle{aa}
\bibliography{reference_list.bib} 

\begin{appendix}
\onecolumn
\section{Slices from the Faraday cube}\label{app:slices}
In this appendix, we present selected slices from the Faraday cube. To make faint structures easier to see, we employed different colour scales for images at different Faraday depths.

\begin{figure}[h!]
\centering
\includegraphics[width=0.8\textwidth]{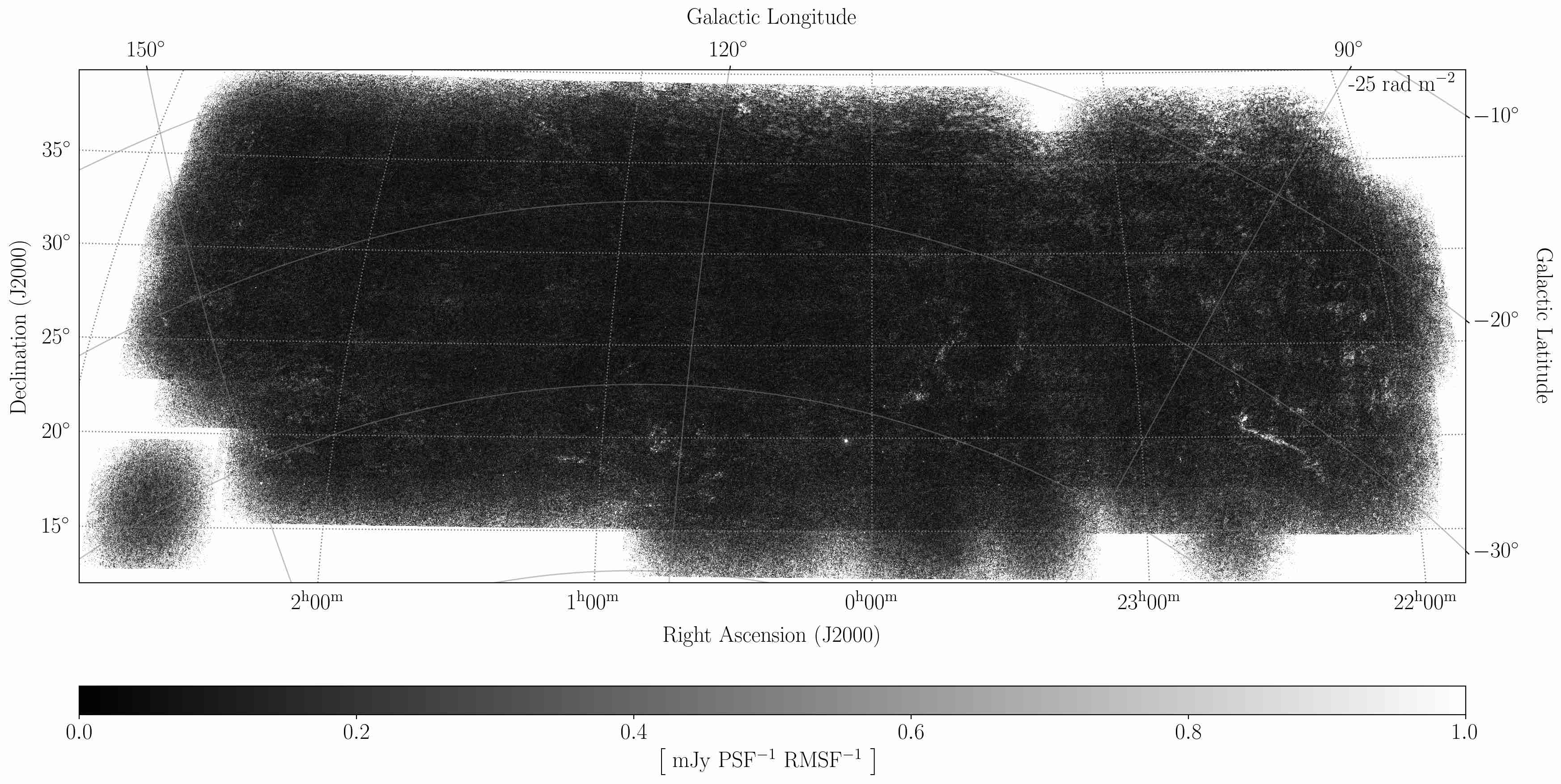}
\includegraphics[width=0.8\textwidth]{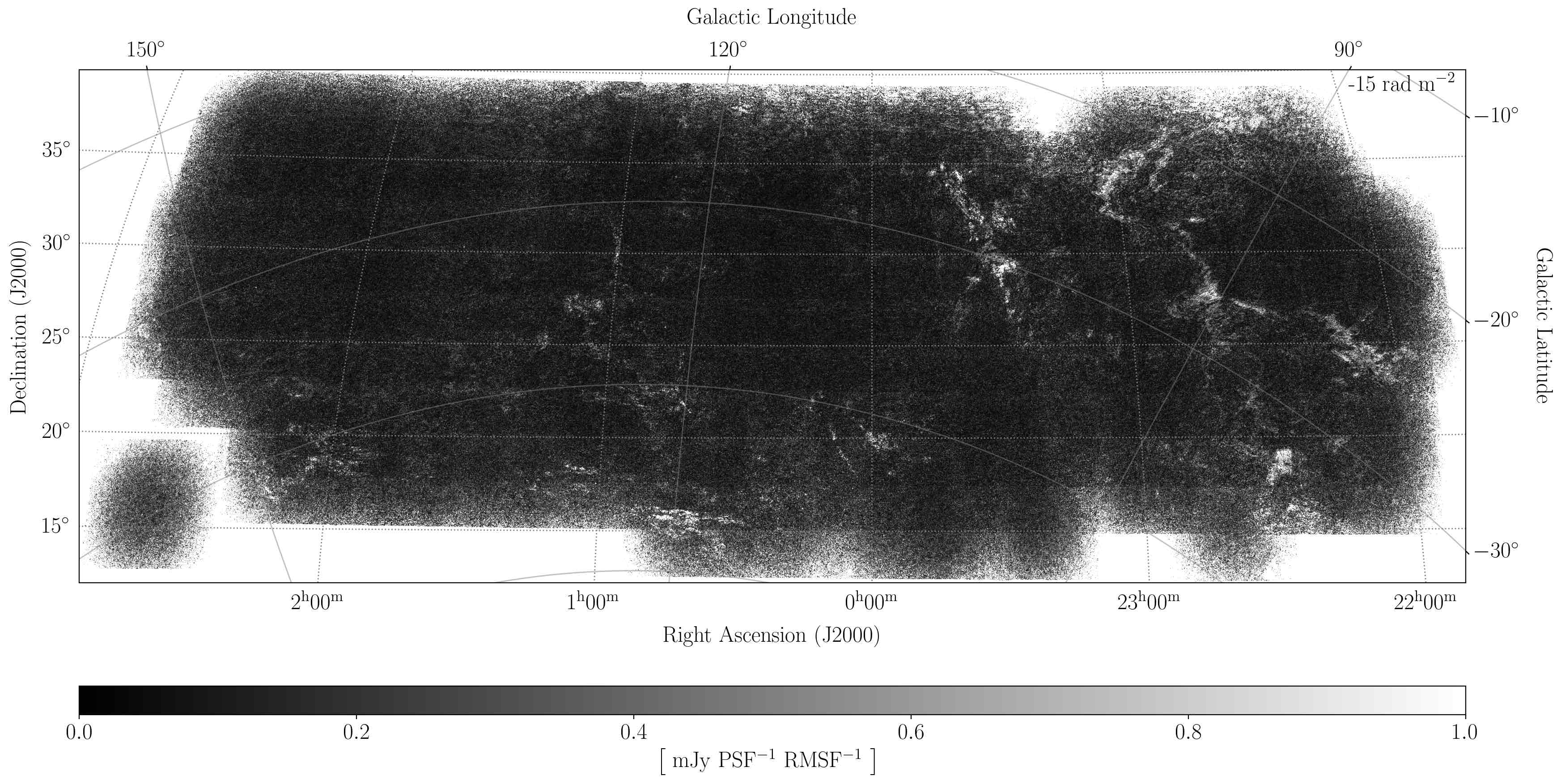}
\includegraphics[width=0.8\textwidth]{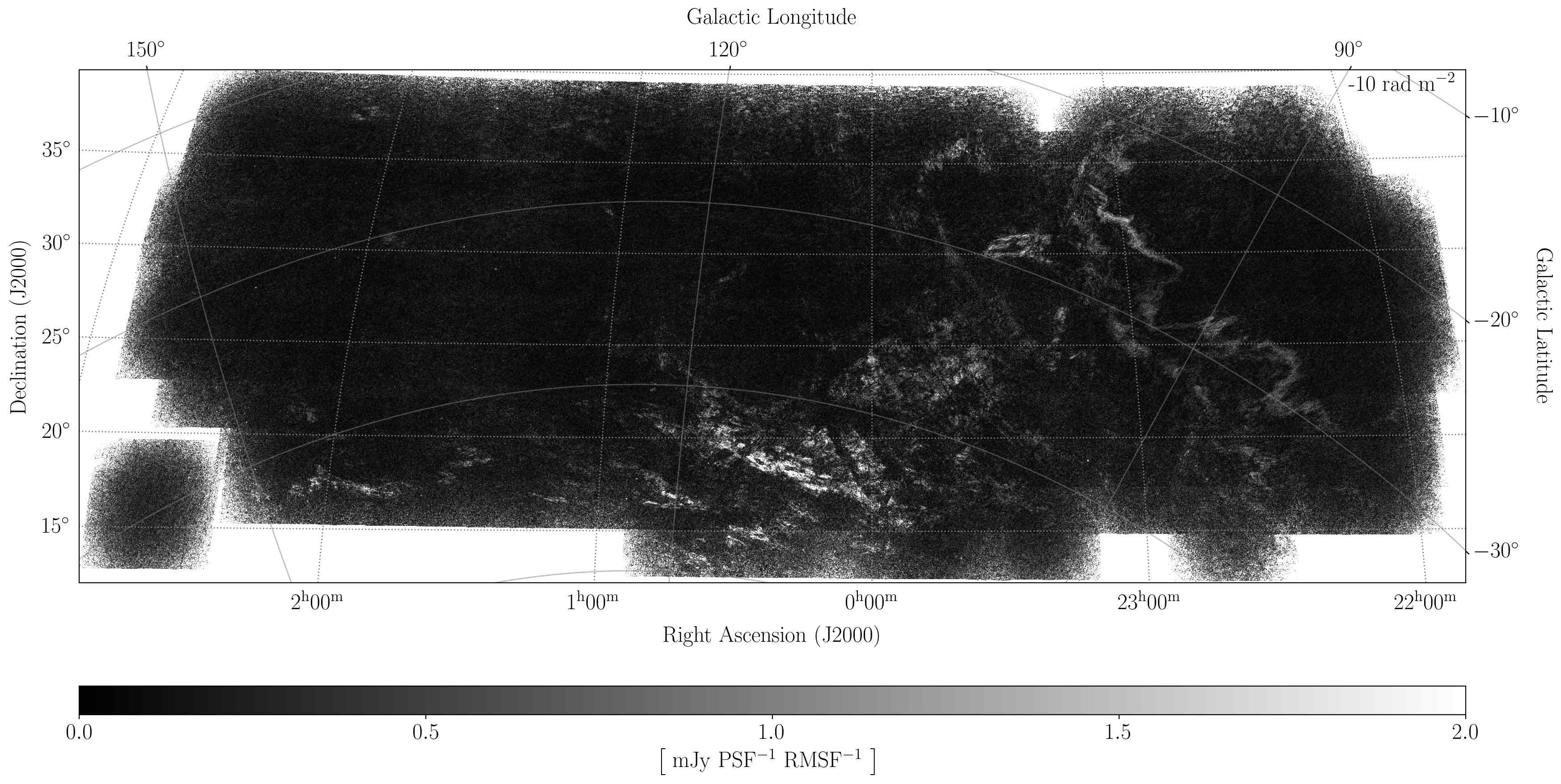}
\caption{Slices from the mosaic Faraday cube at Faraday depths $-25, -15, -10~\mathrm{rad~m^{-2}}$.}
\label{slices1}
\end{figure}

\begin{figure}
\centering
\includegraphics[width=0.8\textwidth]{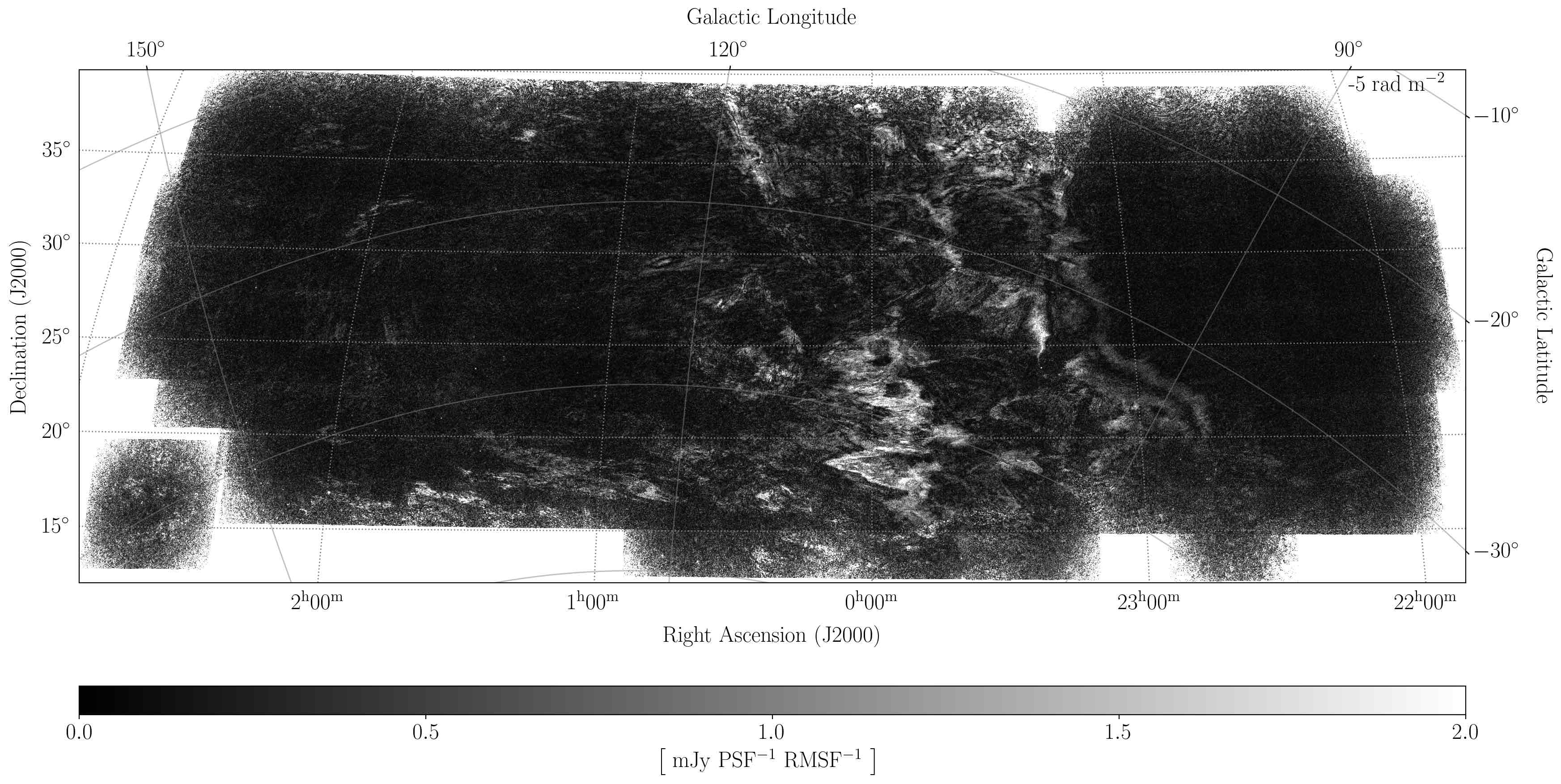}
\includegraphics[width=0.8\textwidth]{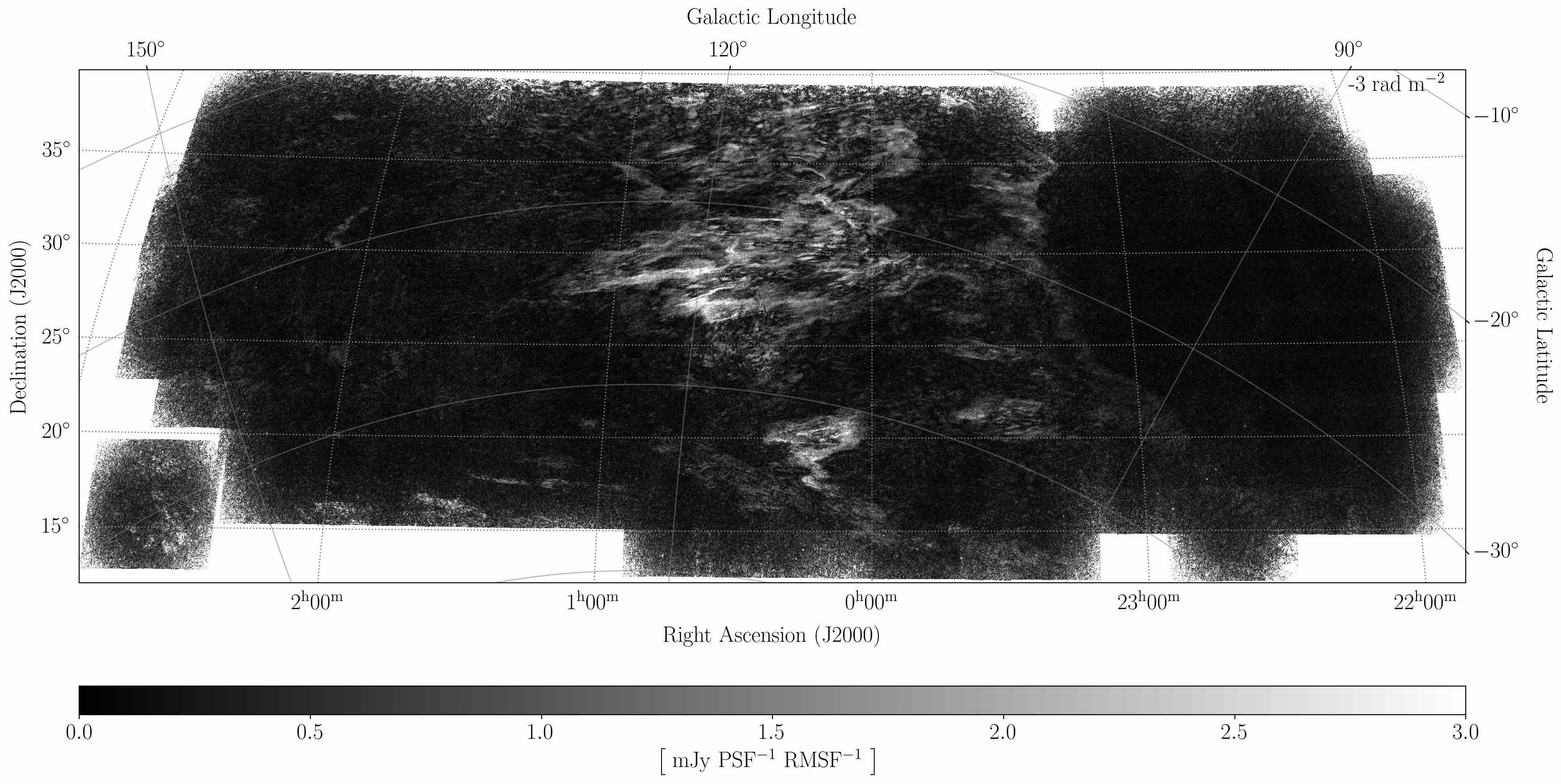}
\includegraphics[width=0.8\textwidth]{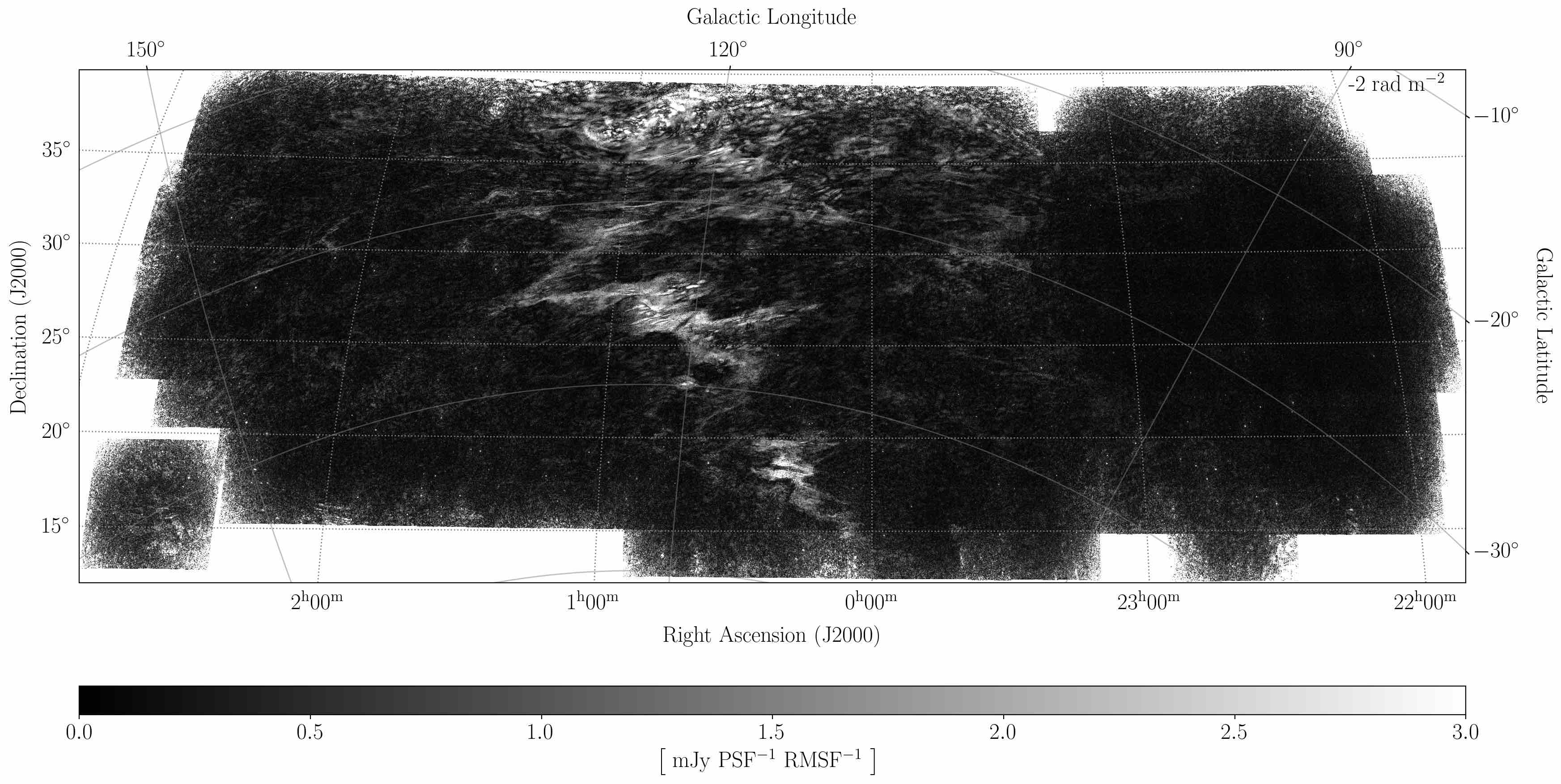}
\caption{Slices from the mosaic Faraday cube at Faraday depths $-5, -3, -2~\mathrm{rad~m^{-2}}$.}
\label{slices2}
\end{figure}

\begin{figure}
\centering
\includegraphics[width=0.8\textwidth]{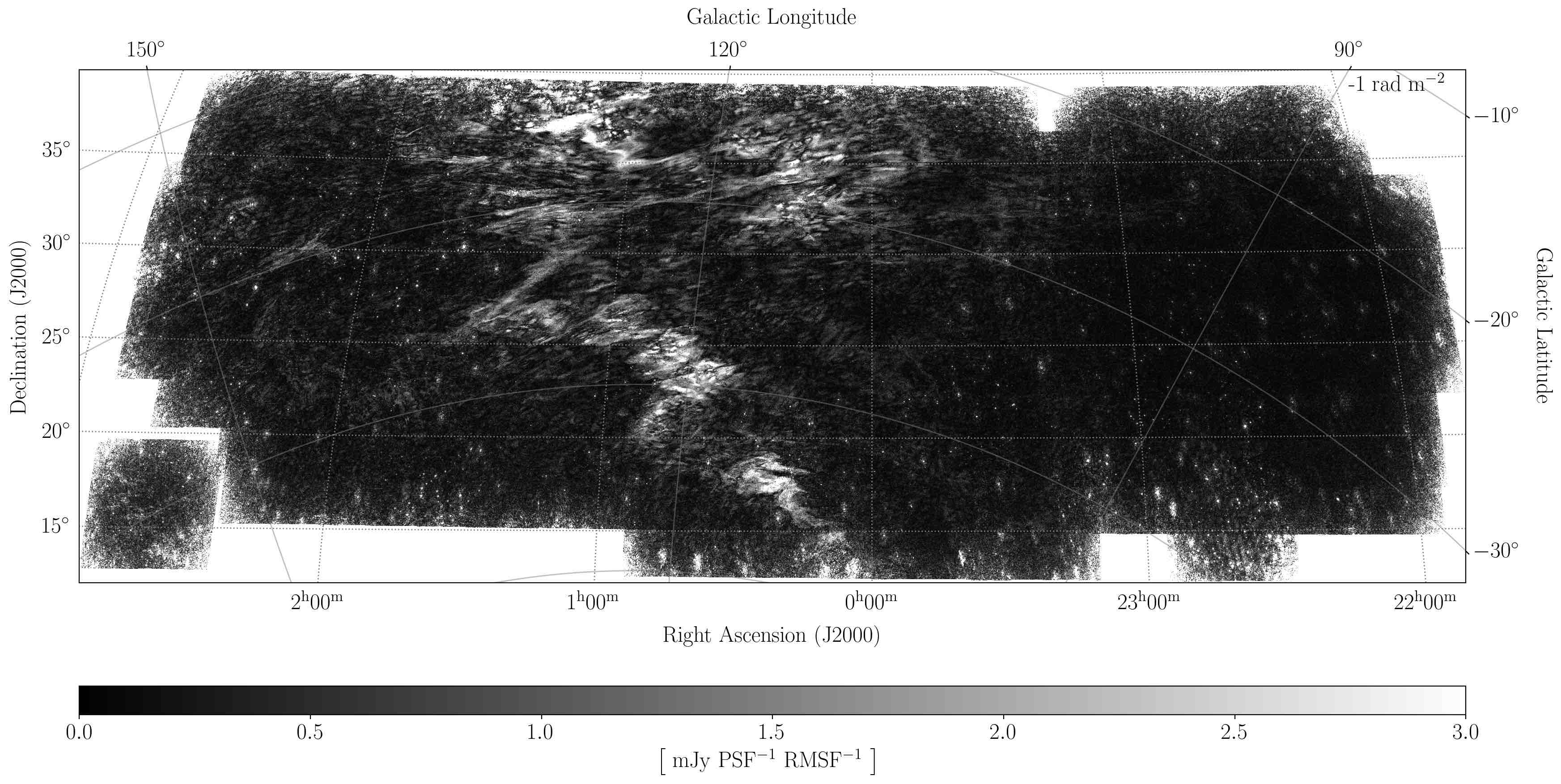}
\includegraphics[width=0.8\textwidth]{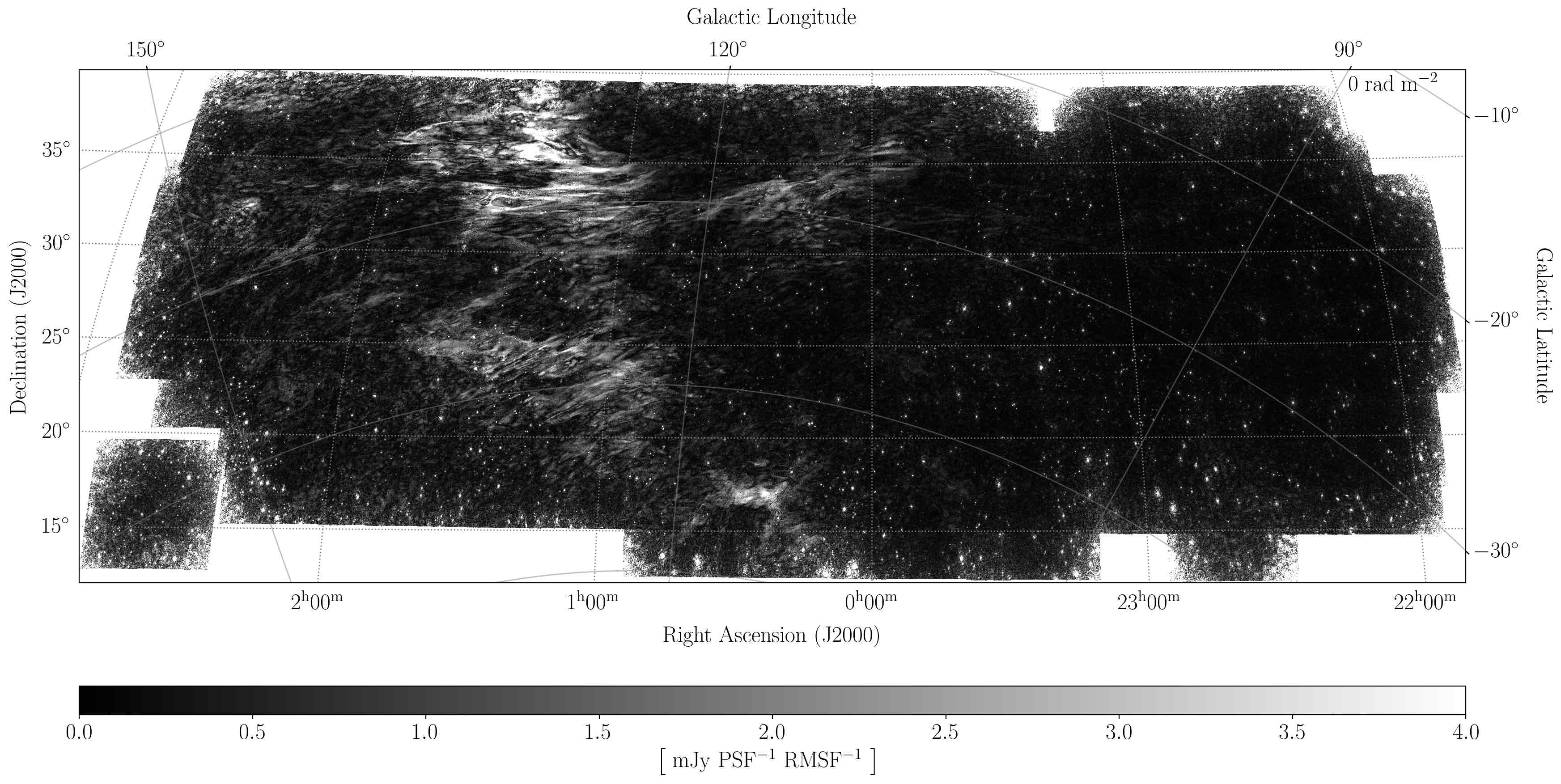}
\includegraphics[width=0.8\textwidth]{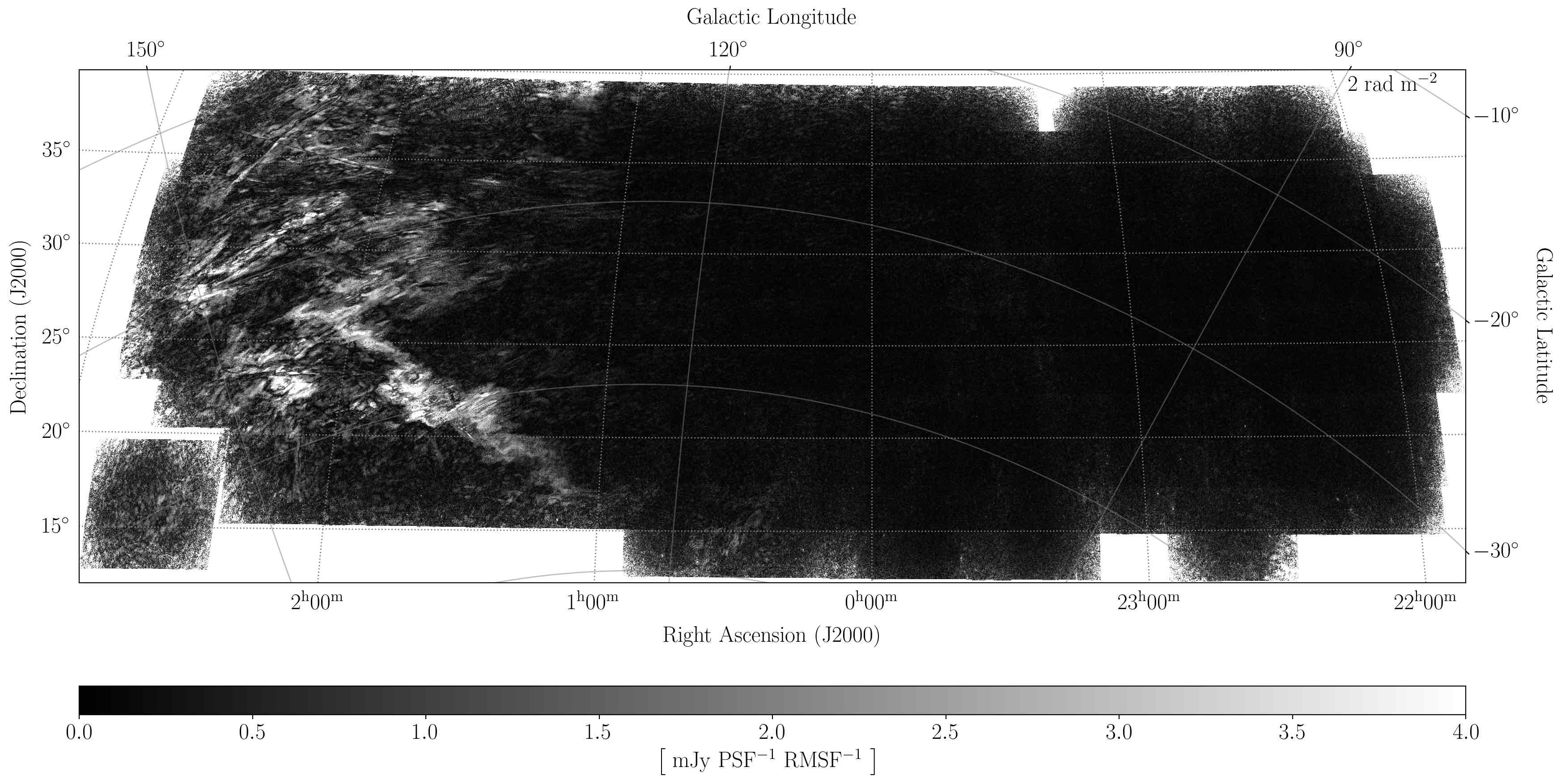}
\caption{Slices from the mosaic Faraday cube at Faraday depths $-1, 0, 2~\mathrm{rad~m^{-2}}$.}\label{slices3}
\end{figure}

\end{appendix}
\end{document}